\documentclass[prd,amsmath,aps,floats,amssymb, floatfix,
  superscriptaddress,nofootinbib,twocolumn,preprintnumbers]{revtex4-2}
\usepackage{tabularx}
\usepackage{amsmath,amssymb,natbib,latexsym,times}

\usepackage{booktabs}
\usepackage{multirow}

\usepackage{graphicx}
\usepackage{dcolumn}
\usepackage{bm}

\usepackage{url}
\usepackage{color}
\usepackage{comment}
\usepackage{mathrsfs}
\usepackage{hyperref}
\usepackage[dvipsnames]{xcolor}
\usepackage{orcidlink}
\usepackage{mathtools}

\newcommand{\mnras}{Monthly Notices of the Royal Astronomical Society}
\newcommand{\physrep}{Physics Reports}
\newcommand{\jcap}{Journal of Cosmology and Astroparticle Physics}
\newcommand{\aj}{The Astronomical Journal}
\renewcommand{\apj}{The Astrophysical Journal}

\newcommand{\aap}{Astronomy and Astrophysics}
\newcommand{\pasj}{Publ.~Astron.~Soc.~Japan}

\newcommand{\araa}{ARA\&A}

\newcommand{\reGauss}{\texttt{reGauss}}

\newcommand{\res}{\mathcal{R}}
\newcommand{\dnnz}{\texttt{dNNz}}

\newcommand{\cmodel}{\texttt{CModel}}

\newcommand{\hmcode}{\textsc{HMCode}}

\begin{document}

\title[]{
Exploring the baryonic effect signature in the Hyper Suprime-Cam Year~3 cosmic shear two-point correlations on small scales: the $S_8$ tension remains present
}

\author{Ryo~Terasawa\orcidlink{0000-0002-1193-623X}}
\email{ryo.terasawa@ipmu.jp}
\affiliation{Kavli Institute for the Physics and Mathematics of the Universe (WPI), The University of Tokyo Institutes for Advanced Study (UTIAS), The University of Tokyo, Chiba 277-8583, Japan}
\affiliation{Department of Physics, The University of Tokyo, Bunkyo, Tokyo 113-0031, Japan}
\affiliation{%
Center for Data-Driven Discovery (CD3), Kavli IPMU (WPI), UTIAS, The University of Tokyo, Kashiwa, Chiba 277-8583, Japan}%

\author{Xiangchong~Li\orcidlink{0000-0003-2880-5102}}
\affiliation{
    McWilliams Center for Cosmology, Department of Physics, Carnegie Mellon
    University, 5000 Forbes Avenue, Pittsburgh, PA 15213, USA
}
\affiliation{
    Kavli Institute for the Physics and Mathematics of the Universe (WPI), The
    University of Tokyo Institutes for Advanced Study (UTIAS), The University
    of Tokyo, Chiba 277-8583, Japan
}

\author{Masahiro~Takada\orcidlink{0000-0002-5578-6472}}
\email{masahiro.takada@ipmu.jp}
\affiliation{Kavli Institute for the Physics and Mathematics of the Universe (WPI), The University of Tokyo Institutes for Advanced Study (UTIAS), The University of Tokyo, Chiba 277-8583, Japan}
\affiliation{%
Center for Data-Driven Discovery (CD3), Kavli IPMU (WPI), UTIAS, The University of Tokyo, Kashiwa, Chiba 277-8583, Japan}%

\author{Takahiro~Nishimichi\orcidlink{0000-0002-9664-0760}}
\affiliation{Department of Astrophysics and Atmospheric Sciences, Faculty of Science, Kyoto Sangyo University, Motoyama, Kamigamo, Kita-ku, Kyoto 603-8555, Japan}
\affiliation{Center for Gravitational Physics and Quantum Information, Yukawa Institute for Theoretical Physics, Kyoto University, Kyoto 606-8502, Japan}
\affiliation{Kavli Institute for the Physics and Mathematics of the Universe
(WPI), The University of Tokyo Institutes for Advanced Study (UTIAS),
The University of Tokyo, Chiba 277-8583, Japan}

\author{Satoshi~Tanaka\orcidlink{0000-0003-2442-8784}}
\affiliation{Center for Gravitational Physics and Quantum Information, Yukawa Institute for Theoretical Physics, Kyoto University, Kyoto 606-8502, Japan}

\author{Sunao~Sugiyama\orcidlink{0000-0003-1153-6735}}
\affiliation{Center for Particle Cosmology, Department of Physics and Astronomy, University of Pennsylvania, Philadelphia, PA 19104, USA}
\affiliation{Kavli Institute for the Physics and Mathematics of the Universe (WPI), The University of Tokyo Institutes for Advanced Study (UTIAS), The University of Tokyo, Chiba 277-8583, Japan}

\author{Toshiki~Kurita\orcidlink{0000-0002-1259-8914}}
\affiliation{Max Planck Institute f\"ur Astrophysik, Karl-Schwarzschild-Str. 1, 85748 Garching, Germany}
\affiliation{Kavli Institute for the Physics and Mathematics of the Universe (WPI), The University of Tokyo Institutes for Advanced Study (UTIAS), The University of Tokyo, Chiba 277-8583, Japan}

\author{Tianqing~Zhang\orcidlink{0000-0002-5596-198X}}
\affiliation{
    McWilliams Center for Cosmology, Department of Physics, Carnegie Mellon
    University, 5000 Forbes Avenue, Pittsburgh, PA 15213, USA
}

\author{Masato~Shirasaki\orcidlink{0000-0002-1706-5797}}
\affiliation{
    National Astronomical Observatory of Japan, National Institutes of Natural
    Sciences, Mitaka, Tokyo 181-8588, Japan
}
\affiliation{
    The Institute of Statistical Mathematics, Tachikawa, Tokyo 190-8562, Japan
}

\author{Ryuichi~Takahashi\orcidlink{0000-0001-6021-0147}}
\affiliation{
    Faculty of Science and Technology, Hirosaki University, 3 Bunkyo-cho,
    Hirosaki, Aomori 036-8561, Japan
}

\author{Hironao~Miyatake\orcidlink{0000-0001-7964-9766}}
\affiliation{
    Kobayashi-Maskawa Institute for the Origin of Particles and the Universe
    (KMI), Nagoya University, Nagoya, 464-8602, Japan
}
\affiliation{
    Institute for Advanced Research, Nagoya University, Nagoya 464-8601, Japan
}
\affiliation{
    Kavli Institute for the Physics and Mathematics of the Universe (WPI), The
    University of Tokyo Institutes for Advanced Study (UTIAS), The University
    of Tokyo, Chiba 277-8583, Japan
}

\author{Surhud~More\orcidlink{0000-0002-2986-2371}}
\affiliation{
    The Inter-University Centre for Astronomy and Astrophysics, Post bag 4,
    Ganeshkhind, Pune 411007, India
}
\affiliation{
    Kavli Institute for the Physics and Mathematics of the Universe (WPI), The
    University of Tokyo Institutes for Advanced Study (UTIAS), The University
    of Tokyo, Chiba 277-8583, Japan
}

\author{Atsushi~J.~Nishizawa\orcidlink{0000-0002-6109-2397}}
\affiliation{
    DX center Gifu Shotoku Gakuen University, Gifu 501-6194, Japan
}
\affiliation{
    Kobayashi-Maskawa Institute for the Origin of Particles and the Universe
    (KMI), Nagoya University, Nagoya, 464-8602, Japan
}

\begin{abstract}
The baryonic feedback effect is considered as a possible solution to the so-called $S_8$ tension indicated in cosmic 
shear cosmology.
The baryonic effect is more significant on smaller scales, and affects the cosmic shear two-point correlation functions (2PCFs) with 
different scale- and redshift-dependencies from those of the cosmological parameters. In this paper, we use the Hyper Suprime-Cam Year~3 (HSC-Y3) data to measure 
the cosmic shear 2PCFs ($\xi_{\pm}$) down to 0.28~arcminutes, taking full advantage of the high number density of source galaxies in the deep HSC data, to explore a possible signature of the baryonic effect. 
While the published HSC analysis used the cosmic shear 
2PCFs on
angular scales, which are sensitive to the matter power spectrum at $k\lesssim 1~h{\rm Mpc}^{-1}$, the 
smaller scale HSC cosmic shear signal allows us to probe the signature of matter power spectrum up to $k\simeq 20~h{\rm Mpc}^{-1}$.
Using the accurate emulator of the nonlinear matter power spectrum, \textsc{DarkEmulator2}, we show that 
the dark matter-only model can provide an acceptable fit to 
the HSC-Y3 2PCFs
down to the smallest scales. 
In other words, we do not find any clear signature of the baryonic effects or do not find a systematic shift 
in the $S_8$ value with the inclusion of the smaller-scale information as would be expected if the baryonic effect is significant. 
Alternatively, we use a flexible 6-parameter model of the baryonic effects, which can lead to both enhancement and suppression in the matter power spectrum compared to the dark matter-only model, to perform the parameter inference of the HSC-Y3 2PCFs. We find that the small-scale HSC data 
allow only a fractional suppression of up to 5 percent in the matter power spectrum at $k\sim 1~h{\rm Mpc}^{-1}$, 
which is not sufficient to reconcile 
the $S_8$ tension.
Finally, we discuss that $\xi_-$ on scales smaller than a few arcminutes 
can be used to monitor characteristic features caused by the baryonic effect if present, while $\xi_{\pm }$ with appropriate scale cuts can be used to constrain the cosmological parameters, simultaneously. 
\end{abstract} 

\maketitle

\section{Introduction}
\label{sec:intro}

The flat $\Lambda$ Cold Dark Matter ($\Lambda$CDM) model 
has been successful in explaining a variety of observations \citep[e.g.,][]{2013PhR...530...87W,2020moco.book.....D}.
{As observational precision has advanced, we find ourselves in the era of precision cosmology, where attention is focused on small deviations between various observations when analyzed within the framework of the flat 
$\Lambda$CDM cosmology model. 
A statistically significant discrepancy after considering known systematic uncertainties could indicate a new physics beyond the flat $\Lambda$CDM cosmology. However, the discrepancy could also arise from unknown systematics in some of the observations or the analyses.
One such discrepancy is known as the $\sigma_8$ or $S_8$ tension. This refers to the consistent lower values of $\sigma_8$ or $S_8$ in $\Lambda$CDM models inferred from large-scale structure probes, which characterizes the clustering amplitude in the present-day universe, compared to those inferred from the \textit{Planck}-2018 CMB measurements~\citep[see][for a recent review]{2022JHEAp..34...49A}.}
Such large-scale structure probes that exhibit the $S_8$ (or $\sigma_8$)
tension include 
cosmological weak lensing \citep[hereafter ``cosmic shear'', e.g.,][]{HSC3_cosmicShearReal,HSC3_cosmicShearFourier,KiDS1000_CS_Asgari2020,DESY3_CS_Secco2022},
joint probe cosmology combining weak lensing and galaxy clustering
\citep[e.g.,][]{HSC3_3x2pt_ss,HSC3_3x2pt_ls,KiDS1000_3x2pt_Heymans2021,DESY3_3x2pt2022}
and redshift-space galaxy clustering \citep[e.g.,][]{2020JCAP...05..042I,2022JCAP...02..008C,2022PhRvD.105h3517K}. 

Among the large-scale structure probes, cosmic shear, which refers to coherent distortions of galaxy images \citep[e.g.,][]{2017grle.book.....D}, provides 
a unique means of probing the distribution of {\it total} 
matter (mainly dark matter) along the line-of-sight direction to the source galaxies.
The lensing distortion is small and can only be measured statistically using more than 
millions of galaxies over a sufficiently wide area of the sky and with various systematic errors under control
\citep{2006MNRAS.366..101H,rev_wlsys_Mandelbaum2017}. 
The ongoing Stage-III surveys, such as the Subaru Hyper Suprime-Cam Survey
\citep[HSC;][]{HSC_hardware_Miyazaki2018,2018PASJ...70S...4A,HSC3_cosmicShearReal,HSC3_cosmicShearFourier,HSC3_3x2pt_ls,HSC3_3x2pt_ss}, the Dark Energy Survey \citep[DES;][]{DESY3_CS_Secco2022,DESY3_CS_Amon2021}, 
and the Kilo-Degree Survey \citep[KiDS;][]{KiDS1000_CS_Asgari2020}, have achieved the precise measurements of 
the cosmic shear two-point correlation functions (2PCFs)
in tomographic redshift bins
to obtain tight constraints on the cosmological parameters of $\Lambda$CDM model.
For cosmic shear cosmology to have sufficient power to constrain cosmological parameters, 
most analyses use the cosmic shear 2PCFs down to small angular scales, which arise from the matter power spectrum  up to 
$k\simeq 1~h{\rm Mpc}^{-1}$ \citep{2005APh....23..369H}, in the nonlinear regime. 
Thus, cosmology inference of cosmic shear 2PCFs 
requires the use of an accurate model template of the nonlinear matter power spectrum for which 
fitting formulae calibrated against $N$-body simulations for different cosmology models
are usually adopted
\citep{halofitT2012,halofit_mead16,halofit_mead21}. 

However, cosmological $N$-body simulations ignore baryonic effects such as star/galaxy formation, supernovae feedback, and AGN feedback in structure formation. The baryonic effect on the matter power spectrum might not be negligible on scales ($k\lesssim 1~h{\rm Mpc}^{-1}$) that are relevant to cosmic shear cosmology. In fact, if the AGN feedback effect, which could push intrahalo gas outside halos into intergalactic space,  
causes a {\it significant} suppression in the matter power spectrum amplitude, it could reconcile the $S_8$ tension \cite{baryon_chen22,baryon_arico23,2022MNRAS.516.5355A,2023MNRAS.525.5554P,baryon_arico23}. 
However, the baryonic effects or more generally galaxy formation and evolution processes involve complicated physics, and are still not possible to model or simulate
from first principles.  
Therefore, simple phenomenological subgrid models 
of the aforementioned baryonic effects
need to be employed to gain insight into the physics of galaxy formation/evolution in the context of cosmological large-scale structure formation
\citep{Illustris_Vogelsberger2014,cowls_LeBrun2014,horizonAgn_Kaviraj2017,2018MNRAS.476.2999M,2018MNRAS.475..676S,IllustrisTNG_Nelson2019,2019MNRAS.486.2827D,FLAMINGO_project}. 
For this reason, 
different cosmological hydrodynamical simulations give different predictions, for example for the matter power spectrum, depending on which subgrid physics model is used \citep{2019OJAp....2E...4C,2020MNRAS.491.2424V,2023ApJ...959..136N}. A standard approach to accounting for the baryonic effect in cosmic shear cosmology is to employ a parametrized model of the baryonic effect on the matter power spectrum and then derive constraints on the cosmological parameters after marginalizing over the parameters of the baryonic effect \citep[e.g.,][]{baryonPCA_Huang2021,HSC3_cosmicShearReal}. 
However, since the model of the baryonic effect is still uncertain, the derived cosmological constraints might not be convincing and depend to some extent on the model of the baryonic effect and the priors used in the analysis. 

Hence, the purpose of this paper is to use an alternative approach to explore a possible signature of the baryonic effect in 
the cosmic shear 2PCFs measured from the Subaru HSC Year~3 (HSC-Y3) data. 
We use an accurate dark matter-only (DM-only) model of the matter power spectrum, instead of the model including the baryonic effect, to perform cosmology inference of the cosmic shear 2PCFs. For the DM-only model, we use the \textsc{DarkEmulator2} of the matter power spectrum, which was calibrated with a suite of $N$-body simulations for different cosmological models, in Tanaka~et al. (in prep.). 
\textsc{DarkEmulator2} was shown to achieve fractional accuracy better than one percent up to 
$k\simeq 100~h{\rm Mpc}^{-1}$ for the range of $\Lambda$CDM models that sufficiently cover the range of
cosmological parameters inferred from the HSC-Y3 cosmic shear results.
We first use the Subaru HSC-Y3 data to 
measure the cosmic shear 2PCFs in tomographic redshift bins down to small angular separations of $0.28$~arcmin, taking full advantage of the high number density of source galaxies. Here the 2PCFs on the smallest angular scales
arise from the matter power spectrum up to 
$k\simeq 20~h{\rm Mpc}^{-1}$.
The baryonic effect is more significant on smaller scales, and affects the cosmic shear 2PCFs with different scale- and redshift-dependences from those of the cosmological parameters. Therefore, if a significant baryonic effect is present in the data, the DM-only model would not give a good fit to the measured 2PCFs with the inclusion of the smaller-scale data points.  
We use the goodness-of-fit to assess the possible failure of the fit for different scale cuts. 
Thus we explore a possible signature of the baryonic effect from the HSC-Y3 cosmic shear 2PCFs by minimizing the uncertainty in the model template or minimizing the impact arising from the use of an uncertain model of the baryonic effect. 
For a more comprehensive discussion, we also 
perform the cosmology inference using a flexible 6-parameter model of the baryonic effect in the matter power spectrum, 
\textsc{HMCode20} \citep{halofit_mead21}, which includes the DM-only model and leads to both enhancement and suppression in the amplitude by varying combinations of the model parameters. We will use the posterior to discuss the range of the matter power spectrum amplitudes as a function of $k$ allowed by the HSC-Y3 cosmic shear 2PCFs.

We organize this paper as follows. In Section~\ref{sec:motivation} we give the motivation and strategy of our study to explore the baryonic effect signature in the HSC-Y3 cosmic shear 2PCFs.
In Section~\ref{sec:data}, we describe the HSC-Y3 data, the new measurement of 2PCFs down to the smallest scales ($0.28$~arcmin), and the covariance matrix.
In Section~\ref{sec:method} we define the different scale cuts used in this paper, and 
describe the model parameters, the priors and the Bayesian inference method. 
In Section~\ref{sec:results}, we show the main results, i.e. the parameter inference of the HSC-Y3 cosmic shear 2PCFs using the DM-only model. 
In Section~\ref{sec:discussion}, we discuss how effects that we ignore in our default method affect the main results.  Section~\ref{sec:conclusion} is devoted to conclusion.

\section{Motivation: baryonic effects on cosmic shear signal}
\label{sec:motivation}

\subsection{Cosmic shear two-point correlation function}
\label{subsec:LCDM}

With the flat-sky approximation, the cosmic shear 2PCFs can be expressed 
in terms of the $E$- and $B$-mode  angular power spectra $C^{E/B}(\ell)$ via the Hankel transform:
\begin{align}
\label{eq:model_hankel}
\xi^{ij}_{+/-}(\theta) = 
\int\!\frac{\ell \mathrm{d}\ell}{2\pi} \,
        \left[ C^{E;ij}(\ell) \pm C^{B;ij}(\ell) \right]J_{0/4}(\theta \ell) ,
\end{align}
where $J_{0/4}$ are the $0$-th/$4$th-order Bessel functions of the first kind.
In our analysis, we use \texttt{FFTLog}
\citep{fftlog2020} to perform 
the Hankel transform.
The superscripts ``$ij$'' denote tomographic bins; e.g. ``$ij$'' means the 2PCF or power spectrum obtained using 
source galaxies in the $i$-th and $j$-th tomographic redshift bins. 

The correlated shapes between different galaxies arise from not only cosmic shear distortion due to foreground structures, 
but also intrinsic alignments (IA) due to the tidal field in the local large-scale structure \citep[e.g., see][and references therein]{2023PhRvD.108h3533K}.
The main purpose of this paper is to assess whether a ``minimal'' theoretical model, i.e. the DM-only structure formation model, can fit the cosmic shear 2PCFs measured from the HSC-Y3 data.
Hence we consider only the cosmic shear contribution, or equivalently ignore the IA contribution when computing the theoretical templates. We will discuss later how our results change when we consider possible contamination of the IA effect. Under this setting, the $E$-mode angular power spectrum in Eq.~(\ref{eq:model_hankel}) is given as
\begin{align}
C^{E;ij}(\ell)=\int^{\chi_H}_0\!\mathrm{d}\chi\frac{q_i(\chi)q_j(\chi)}{\chi^2}P_{\rm m}\!\!\left(k=\frac{\ell+1/2}{\chi};z(\chi)\right) ,
\label{eq:cl_EE}
\end{align}
where $P_{\rm m}(k;z)$ is the matter power spectrum at redshift $z$, 
$\chi(z)$ is the radial comoving distance up to redshift $z$, $\chi_H$ is the distance to the horizon, and 
$q_i(\chi)$ is the lensing efficiency function for source galaxies in the $i$-th redshift bin, defined as
\begin{align}
q_i(\chi)=\frac{3}{2}\Omega_{\rm m}H_0^2\frac{\chi}{a(\chi)}\int_\chi^{\chi_H}\!\mathrm{d}\chi'~n_i(\chi')\frac{\chi'-\chi}{\chi'},
\end{align}
where $\Omega_{\rm m}$ is the present-day density parameter of matter, $H_0$ is the present-day Hubble constant ($H_0=100h~{\rm km}~{\rm s}^{-1}~{\rm Mpc}^{-1}$), $a$ is the scale factor, and $n_i(\chi)$ is the normalized redshift distribution of galaxies in the $i$-th source redshift bin. 
Note that $\chi$ is given as a function of redshift as $\chi=\chi(z)$, and $z$ is given by the 
inverse function as $z=z(\chi)$.

\subsection{Baryonic effect on cosmic shear 2PCFs}
\label{subsec:model_baryon}

\begin{figure*}
    \includegraphics[width=2.1\columnwidth]{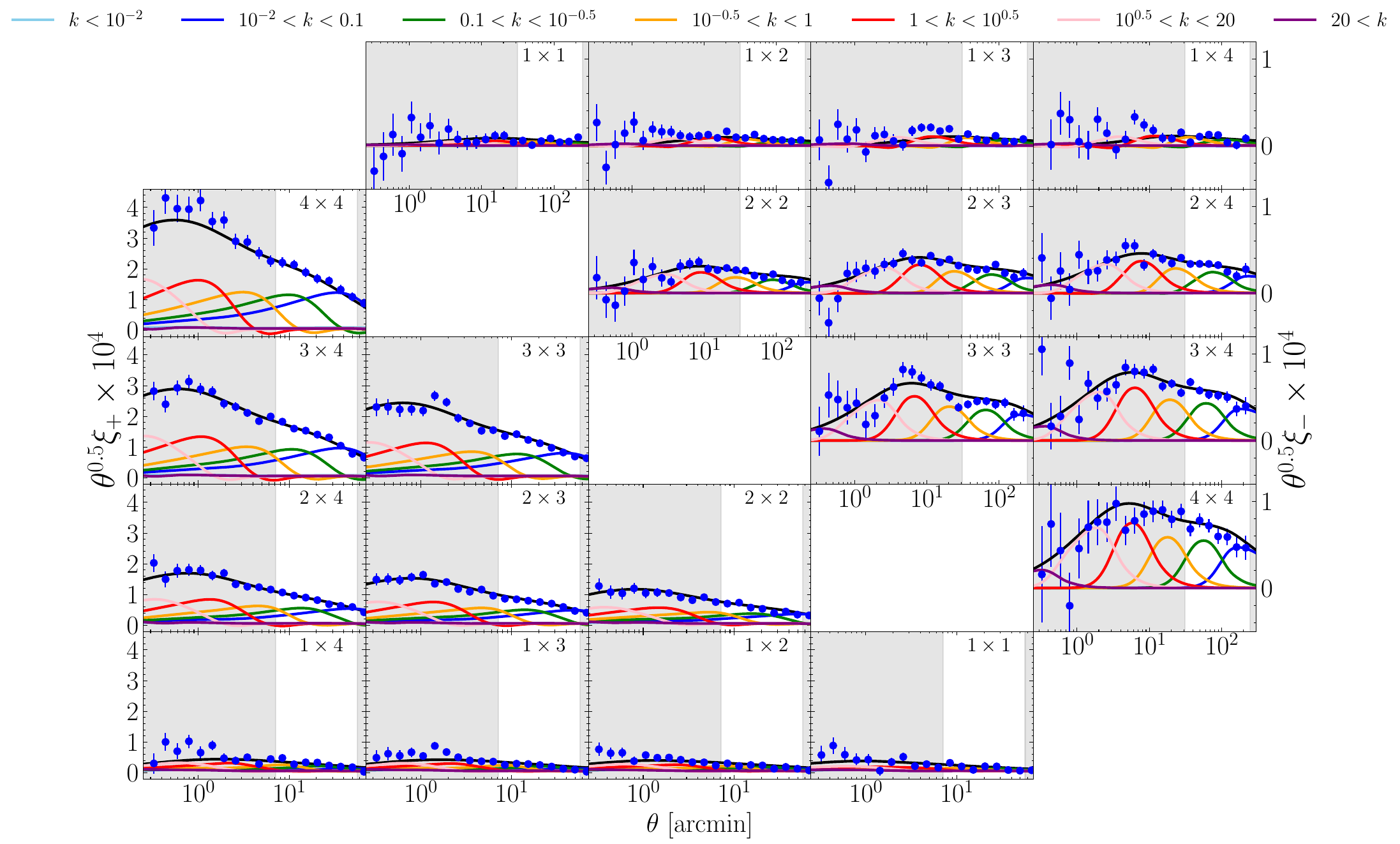}
    \caption{Cosmic shear two-point correlation functions: the lower-left diagonal panels are for $\xi_+(\theta)$, while 
    the upper-right diagonal panels are for $\xi_-(\theta)$. Different panels show the auto- and cross-2PCFs for galaxies in two tomographic redshift bins; for instance, $\xi_{\pm}$ for ``$3\times 4$'' are the 2PCFs of source galaxies 
    in the 3rd- and 4th- redshift bins. Note that the plotted $y$-range in the different-row panels of $\xi_+$ or $\xi_-$ is the same. 
    For the sake of illustration, 
    we plot $\theta^{0.5}\times \xi_{\pm}$. 
    The black line in each panel shows the model prediction for the fiducial 
    $\Lambda$CDM model. The other-color lines show the model predictions obtained by using the Hankel transform 
    of $P_{\rm m}(k)$ over a finite range of $k$ as denoted in the top legend: e.g., $10^{-0.5}<k<1$ denotes the 
    range of $[10^{-0.5},1]~h{\rm Mpc}^{-1}$. For comparison, the data points with error bars denote the measurements from the HSC Year~3 data. The unshaded $x$-axis region denotes the {\it fiducial} scale cuts of $\theta$, which are used for the cosmology inference of the HSC Year~3 measurements in \citet[][hereafter Li+HSCY3]{HSC3_cosmicShearReal}.
    }
    \label{fig:xipm_k-modes}
\end{figure*}
\begin{figure}
 \includegraphics[width=\columnwidth]{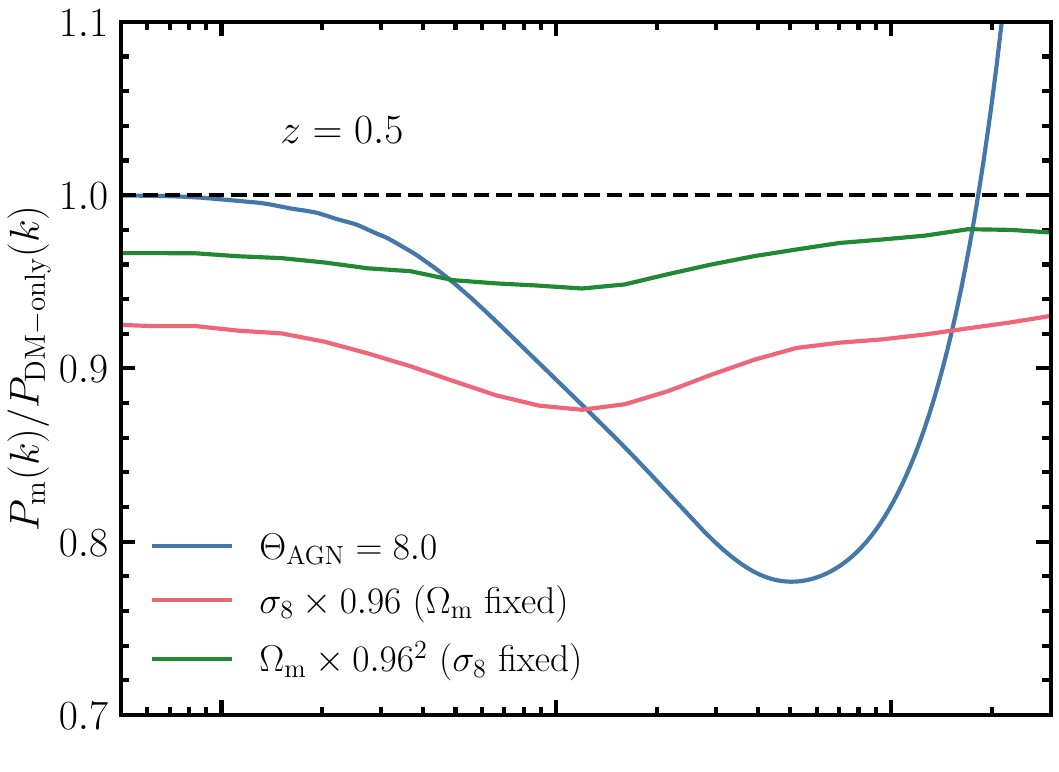}
 \includegraphics[width=\columnwidth]{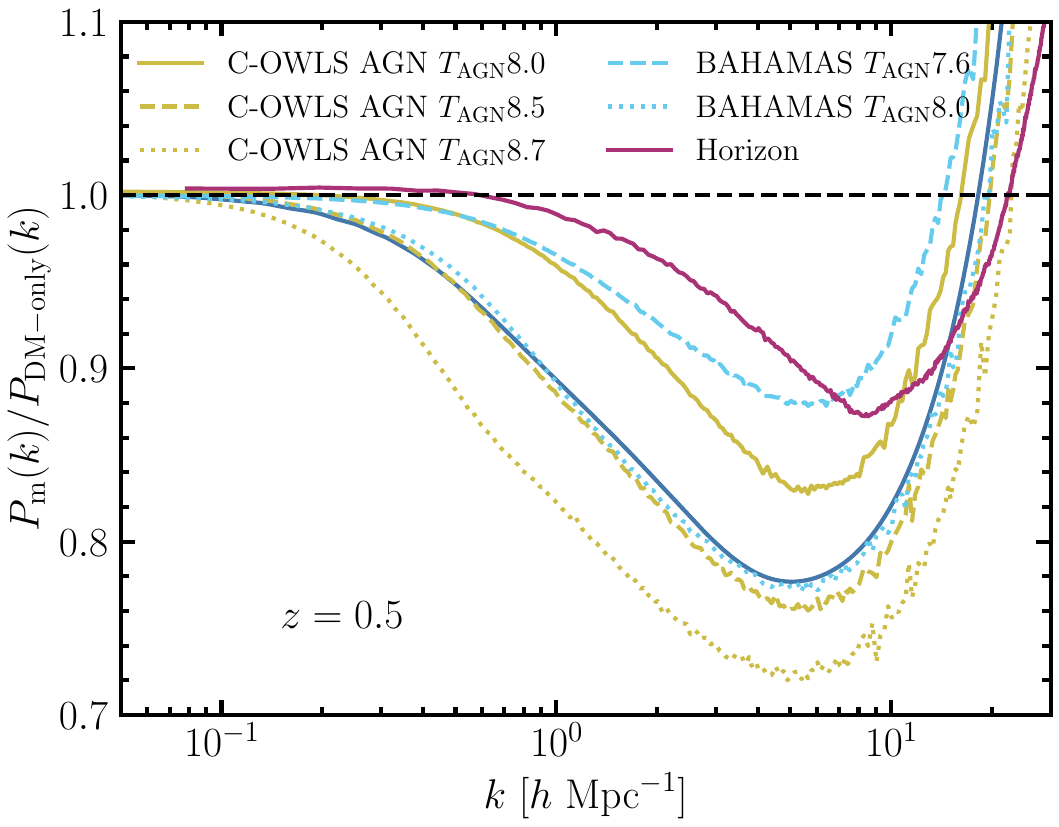}
 \caption{{\em Upper panel}: fractional changes in the matter power spectrum $P_{\rm m}(k)$ at $z=0.5$ due to the baryonic effects 
 with AGN feedback parameter of $\Theta_{\rm AGN}\equiv \log_{10}(T_{\rm AGN}/{\rm K})=8.0$ and changes 
 in the cosmological parameter $\Omega_{\rm m}$ or $\sigma_8$, respectively. Here we consider a flat $\Lambda$CDM model, which is consistent with the {\it Planck} cosmology, and the change in $\Omega_{\rm m}$ and $\sigma_8$ corresponds to 
 4\% fractional change in $S_8\equiv \sigma_8(\Omega_{\rm m}/0.3)^{0.5}$.
 We used \textsc{HMCode20} \citep{halofit_mead21} to compute the matter power spectrum with the AGN feedback effect. 
 {\em Lower panel}: Similar plot, but the figure shows fractional changes in the matter power spectrum compared to the DM-only model, found in
 the hydrodynamical simulations. We use the power spectrum library in \citet{2020MNRAS.491.2424V}.
 Here we show the results from \textsc{COSMO-OWLS} \citep{cowls_LeBrun2014}, 
 \textsc{BAHAMAS} \citep{2018MNRAS.476.2999M}, and \textsc{Horizon} \citep{horizonAgn_Kaviraj2017}.
}
 \label{fig:Pk_ratio}
\end{figure}
\begin{figure*}
    \includegraphics[width=2.1\columnwidth]{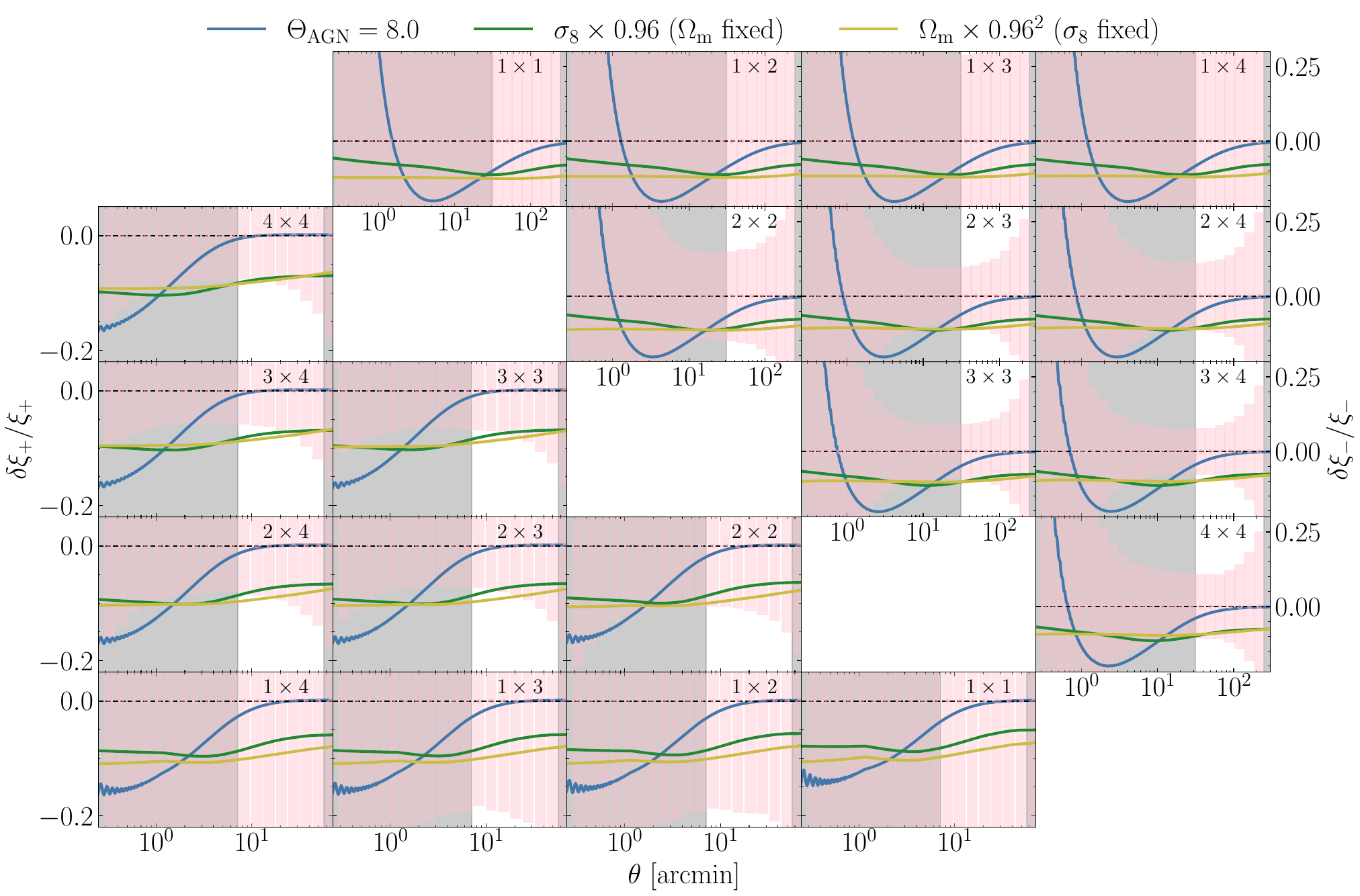}
    \caption{Similarly to Fig.~\ref{fig:Pk_ratio}, we show fractional changes in the cosmic shear 2PCFs, $\xi_{\pm}$, in the four tomographic redshift bins (see Fig.~\ref{fig:xipm_k-modes}). The magenta shaded region in each 
    separation bin shows the $1\sigma$ statistical error calculated from the diagonal component of the covariance matrix for the HSC-Y3 data used in Li+HSCY3. Note that the errors between the neighboring bins can be highly correlated, especially on large scales. The gray shaded region denotes the scale cut used in 
    Li+HSCY3; the data in the unshaded region was used for the cosmology inference. 
    }
    \label{fig:xipm}
\end{figure*}
The fundamental quantity needed to obtain the theoretical template of cosmic shear 2PCFs is the 3D matter power spectrum, 
$P_{\rm m}(k;z)$ (see Eqs.~\ref{eq:model_hankel} and \ref{eq:cl_EE}). The Hankel transform involves a conversion from the 3D wavenumber ($k$) to the angular scales ($\theta$). The model requires a computation 
of $P_{\rm m}(k)$ for a given $\Lambda$CDM model; whilst the linear theory is valid to accurately model $P_{\rm m}$
on large scales ($k\lesssim 0.1~h{\rm Mpc}^{-1}$), on small scales (large $k$) we need to include the nonlinear clustering effect on $P_{\rm m}$, which can be accurately calibrated using $N$-body simulations.

Fig.~\ref{fig:xipm_k-modes} shows the cosmic shear 2PCFs, $\xi_\pm(\theta)$, for the best-fit $\Lambda$CDM model in \citet[][hereafter Li+HSCY3]{HSC3_cosmicShearReal}.
The 2PCFs of source galaxies in higher redshift bins have greater amplitudes because of higher lensing efficiency and more intervening matter contributions up to galaxies in the redshift bins from us. 
The different lines in each panel show 
which range of wavenmubers in $P_{\rm m}(k)$ contributes to 
$\xi_\pm(\theta)$ at each separation angle ($\theta$). 
To be more quantitative, if we focus on the plot of $\xi_{+}$ for the $4\times 4$ source redshift bins, 
the lines using $P_{\rm m}(k)$ with $k$-cuts of 
$0.1<k<10^{-0.5}$, $10^{-0.5}<k<1$, $1<k<10^{0.5}$ and $10^{0.5}<k<20$ 
(all in the units of $[h{\rm Mpc}^{-1}]$) have most of the power up to 
 $\theta\simeq 40$, 13.8, 4.3 and 1.5~arcmin, respectively. On the other hand, the figure shows that $\xi_{-}$ have the power up to about a factor of 4 larger separation than in $\xi_+$:
\begin{align}
\theta_-(<k)\sim 4\theta_+(<k). 
\label{eq:theta-4theta+}
\end{align}
Thus, for a fixed separation of $\theta$, $\xi_-$ is more sensitive to $P_{\rm m}(k)$ at higher $k$ (smaller length scales)
\citep[also see Fig.~23 in Ref.][for the similar discussion]{cosmicShear_HSC1_Hamana2019}.
For the fiducial scale cuts (unshaded range in the $x$-axis) that were used for the cosmology inference in Li+HSCY3, the matter power spectrum at $k\lesssim 1~h{\rm Mpc}^{-1}$ gives the majority of the contributions over the range of angular scales 
\citep[aslo see][]{2005APh....23..369H}. In other words, the cosmic shear 2PCFs at the smaller angular scales
allow us to probe the power spectrum of {\it total} matter on the smaller (higher $k$) scales that are more affected by baryonic effects, 
if those are present in the data. 

In Fig.~\ref{fig:Pk_ratio}, we show how an example model of the baryonic feedback effect changes $P_{\rm m}(k)$ at $z=0.5$ as a function of 
$k$, relative to the DM-only model prediction of $P_{\rm m}(k)$. Here we use \hmcode20 to model the AGN feedback effect on the matter power spectrum, where a phenomenological parameter, $\Theta_{\rm AGN} \equiv \log_{10}(T_{\rm AGN}/{\rm K})$, is used to mimic the AGN-driven feedback effect on $P_{\rm m}(k)$
in hydrodynamical simulations.
Here we consider the case for $\Theta_{\rm AGN}=8.0$, which can reproduce $P(k)$ measured from 
the corresponding realization of the \textsc{BAHAMAS} simulations 
\citep{2018MNRAS.476.2999M} \citep[also see][]{FLAMINGO_project}. The model causes about 10\% suppression 
at $k\simeq 1~h{\rm Mpc}^{-1}$ and the greater suppression in the $P_{\rm m}$ amplitude at the larger $k$. For even larger $k$, $k \gtrsim {\rm a~ few} \times 10~h{\rm Mpc}^{-1}$, the model turns to cause enhancement.
For comparison, we also show the dependence of $P_{\rm m}$ on cosmological parameters, $\Omega_{\rm m}$ 
and $\sigma_{8}$, to which the cosmic shear 2PCFs are most sensitive. The figure shows that the cosmological parameters change the power spectrum on all scales, with different $k$-dependences compared to the baryonic effect.

In Fig.~\ref{fig:xipm} we show how the baryonic effect and the cosmological parameters affect the cosmic shear 2PCFs in different tomographic redshift bins.
As can be seen (also see Fig.~\ref{fig:xipm_k-modes}), the baryonic effect with $\Theta_{\rm AGN}=8.0$ causes a more significant change in 
the 2PCFs on small scales, compared to the changes due to the cosmological parameters.
For the scale cut used in Li+HSCY3, 
the fractional changes in $\xi_\pm$ by the baryonic effect are smaller than the statistical errors of HSC Year~3 data, leading to a consequence that the scale cut appears to avoid the impact of the baryonic effect on the cosmological parameters, if 
the magnitude of the baryonic effect is at the level of $\Theta_{\rm AGN}=8.0$. 
Nevertheless, it is also interesting to note that $\xi_-$ is more affected by the baryonic effect 
than $\xi_+$ at the same angular separation. 
We will use these different features to test a possible impact of the baryonic effect in the HSC Year~3 data.

\subsection{Analysis strategy}
\label{ssec:strategy}

All the analyses presented in this paper were done after the unblinding (see Li+HSCY3 for details of the blinding HSC-Y3 cosmology analysis). 
The main purpose of this paper is to explore a possible signature of the baryonic effect in the HSC-Y3 cosmic shear 2PCFs. 
Based on this motivation, we will employ analysis setups that allow us to 
illuminate the possible baryon signature, if it is present in the HSC data.

\section{Data}
\label{sec:data}

In this section, we briefly introduce the HSC-Y3 data used in  the cosmic shear
analysis. The data is based on the S19A internal data release, which was
released in September 2019 and was acquired between March 2014 and April 2019.
For the details of the data and catalog used in this paper, please see 
\citet{HSC3_catalog_Li2021} 
for the shear catalog, \citet{HSC3_photoz_Nishizawa2020} and 
\citet{HSC3_photoz_Rau2022}
for the photometric redshift methods, \citet{2023MNRAS.525.2441Z}
for the method to characterize the effect of residual PSF systematics, 
\citet{HSC1_mock_Shirasaki2019} for the mock catalogs of HSC data, 
and Li+HSCY3 for the measurements.

\subsection{Weak-lensing Shear Catalog}
\label{subsec:data_shear}

\subsubsection{Basic characterization}
\label{subsubsec:data_shear_basic}

The original HSC-Y3 shape catalog \citep{HSC3_catalog_Li2021} contains more
than 35 million source galaxies covering $433~\mathrm{deg}^2$ of the northern
sky. The galaxy sample is conservatively selected for the weak-lensing science
with a magnitude cut on extinction-corrected \cmodel{} magnitude at $i <
24.5$\,, a \cmodel{} signal-to-noise ratio (SNR) cut at $\mathrm{SNR}>10$ and a
\reGauss{} resolution cut at $R_2 > 0.3$ \citep{HSC3_catalog_Li2021}.

After the production of the shear catalog, a few additional cuts are applied to
improve the data quality. In particular, we follow
\citep{KiDS450_cs_Hildebrandt2017} to remove objects with extremely large
$i$-band ellipticity which are potentially unresolved binary stars. To be more specific, we remove objects with large ellipticity, $|e| > 0.8$ and
$i$-band determinant radius $r_\mathrm{det} < 10^{- 0.1 r + 1.8}$ arcsec (where
$r$ in the exponent is the r-band magnitude), amounting to $0.46\%$ of the
galaxy sample \citep{HSC3_catalog_Li2021}.

In addition, we remove a region in GAMA09H with $132.5 < \mathrm{RA}<
140$~[deg], $1.6<\mathrm{Dec}<5$~[deg],  containing an area of
$\sim$$20~\mathrm{deg}^2$. This region has the very good seeing size of
$\sim$$0.4~\mathrm{arcsec}$, but it has a smaller number of single-frame
exposures contributing to the coadded images. In addition, we find significant
PSF fourth moment modeling errors in this region \citep{2023MNRAS.525.2441Z}.
We find that
including galaxy shapes in this region causes significant $B$-modes in 2PCFs at
high redshifts and large scales.

Additionally, a number of galaxies are found to have secondary solutions at
very high redshifts in their estimated photo-$z$ posterior distributions, due
to redshift template degeneracies. These secondary solutions are outside the
redshift coverage of our CAMIRA-LRG sample \citep{CAMIRA_HSC_Oguri2018} making
it difficult to calibrate with the cross-correlation technique
\citep{HSC3_photoz_Rau2022}. 

After these cuts, the final shear catalog contains 25 million galaxies covering
416 deg$^2$ of the northern sky. The catalog is split into six subfields: XMM,
GAMA09H, WIDE12H, GAMA15H, VVDS and HECTOMAP. The area and effective galaxy
number densities, $n_\text{eff}$ \citep[as defined in
Ref.][]{WLsurvey_neffective_Chang2013}, in different redshift bins of the
subfields are summarized in Table~I of Li+HSCY3.
The number density maps
for six subfields are shown in Fig.~1 of the same paper.
The effective standard deviation of the error on the per-component shear per galaxy is
$\sigma_\gamma = 0.236$\,.

\subsubsection{Galaxy shear}
\label{ssubsec:shearcat_galshapes}

The HSC-Y3 shear catalog contains galaxy shapes, estimated with the
re-Gaussianization (\reGauss{}) PSF correction method \citep{Regaussianization}
from the HSC $i$-band wide-field coadded images \citep{HSC1_pipeline}. The
\reGauss{} estimator measures the two components of galaxy ellipticity.
The estimated shear for the galaxy ensemble after calibration is
\begin{equation}\label{eq:shear_ensemble_pre}
    \hat{\gamma}_{\alpha}=\frac{\sum_i w_i e_{\alpha;i}}
        {2\, \mathcal{R} (1+\hat{m})\sum_i w_i}
    -\frac{\hat{c}_\alpha}{1+\hat{m}}\,,
\end{equation}
where $\alpha = 1,2$\,, the $e_{\alpha;i}$ is the galaxy ellipticity component for the $i$-th galaxy, 
and $\hat{m}$ and $\hat{c}$ are the calibration factors of multiplicative and additive biases, respectively, that are
given on individual galaxy basis using the image simulations \citep{HSC3_catalog_Li2021,HSC3_cosmicShearReal}.
The galaxy shape weight $w_i$ is defined as
\begin{equation}
    w_{i} =\frac{1}{\sigma_{e;i}^2+e_{\text{rms};i}^2},
\end{equation}
where $e_{\text{rms};i}$ is the root-mean-square ($\texttt{RMS}$) of the
intrinsic ellipticity per component for the $i$-th galaxy. $e_{\rm{rms}}$ and
$\sigma_e$ are modeled and estimated for each galaxy using the image
simulations. ${\cal R}$ is the shear responsivity \citep{2002AJ....123..583B} for the galaxy
population, defined as
\begin{equation}\label{eq:response}
    \res=1-\frac{\sum_i w_i e^2_{\mathrm{rms};i}}{\sum_i w_i}\,.
\end{equation}
We will also take into account the selection bias following the method in Section~IIA3 of 
Li+HSCY3.

\subsubsection{Tomographic redshift bins}
\label{sssec:photoz}

In this paper we use the photometric redshift method, referred to as \dnnz, to define the tomographic 
redshift bins \citep[see][for details]{HSC3_photoz_Nishizawa2020}. 
\dnnz~ is a photo-$z$ conditional density estimation algorithm based on a neural network. The code
uses \cmodel~ fluxes, convolved fluxes, PSF fluxes, galaxy sizes, and galaxy shapes for the training. 
The photo-$z$ conditional density is constructed with 100 nodes in the output layer, and each node 
represents a redshift histogram bin spanning from $z=0$ to $z=7$.

We divide the galaxies in the shape catalog into four tomographic redshift bins by selecting 
galaxies using the best-fit (point) estimate of \dnnz~output for individual galaxies \citep[also see][for details]{HSC3_cosmicShearReal} -- $(0.3,0.6]$, $(0.6,0.9]$, $(0.9,1.2]$ and $(1.2,1.5]$. However, we found that about 
$31\%$ and $8\%$ galaxies in the first and second redshift bins, respectively, have double peaks in the \dnnz~photo-$z$ probability density function, where the secondary peaks correspond to possible outliers at $z\gtrsim 3$.
We remove these galaxies from our source galaxy sample by imposing the additional cuts 
described in Section~IIC of Li+HSCY3.
The first, second, third, and fourth redshift bin contains 
about 6.7, 7.6, 6.0, and 2.2 million galaxies, respectively (see Table~I in Li+HSCY3). 
The intrinsic redshift distributions of galaxies at $z\lesssim 1.2$ residing
in the first and second bins and partially the third bin, denoted $n_i(z)$, were 
reconstructed based on the 
calibration method \citep{HSC3_photoz_Rau2022}
using the cross-correlations of the HSC source galaxies with the red-sequence galaxies, CAMIRA
LRGs \citep{CAMIRA_HSC_Oguri2018}, where the latter LRGs have relatively accurate photo-$z$ estimates. 

Nevertheless, we admit the fact that the redshift distribution of HSC galaxies at $z\gtrsim 1.2$ cannot be calibrated by the cross-correlation method due to the lack of CAMIRA LRGs at such high redshifts.
To perform robust cosmology analysis, we follow the same method in Li+HSCY3 
\citep[also see][]{HSC3_cosmicShearFourier}; 
we adopt the nuisance parameters to model possible residual systematic errors in the mean of source redshifts 
for galaxies in the third and fourth redshift bins.
For the possible residual redshift errors for galaxies at $z\gtrsim 1$, also see 
the companion papers of the HSC Year~3 data, \citet{HSC3_3x2pt_ss} and \citet{HSC3_3x2pt_ls}, for details, where 
the residual redshift errors were calibrated by using a different calibration sample from the CAMIRA LRGs,
which is a sample of the spectroscopic SDSS galaxies. 
More explicitly, we model the underlying true redshift distribution of source galaxies by
shifting the inferred redshift distribution 
of galaxies as
\begin{align}
n_i(z) \rightarrow n_i(z+\Delta z_i),
\label{eq:dz_def}
\end{align}
where $i=1, \dots, 4$  and $\Delta z_i$ is a nuisance parameter that is included in the cosmology inference. 
Note that $\Delta z_i<0$ means that the mean redshift of the true distribution is {\it higher} than that inferred from the photo-$z$ method. 
In our method, we assume that the shape of the redshift distribution is the same as that inferred from the 
photo-$z$ calibration method, but the redshift shift in the source redshifts is modeled by $\Delta z_i$. 
$\Delta z_i$ can model changes in relative amplitudes of the tomographic 2PCFs, while a change in the shape of 
$n_i(z)$ alters the shape (scale-dependence) of 2PCFs.
Thus, although our method is a simplified method, we conclude that our treatment is sufficient for the current statistical precision of the HSC-Y3 cosmic shear 2PCFs.

\subsection{New measurements of cosmic shear 2PCFs}
\label{ssec:2pcfs_measurements}

Since our focus is on exploring the baryonic effect of the cosmic shear signals on smaller scales than done 
in Li+HSCY3, 
we re-measure the cosmic shear 2PCFs from the HSC-Y3 shape catalog \citep{HSC3_catalog_Li2021}. We can estimate the 2PCFs as
\begin{align}
\hat{\xi}_{\pm}(\theta)&=
\frac{\sum_{i,j}w_i w_j \gamma_+({\bf n}_i)
\gamma_+({\bf n}_j)
}{\sum_{i,j}w_iw_j}\nonumber\\
&\hspace{1em}\pm
\frac{\sum_{i,j}w_i w_j \gamma_\times({\bf n}_i)
\gamma_\times({\bf n}_j)
}{\sum_{i,j}w_iw_j},
\end{align}
where $i$, $j$ are the $i$- and $j$-th indices of two galaxies, ${\bf n}_{i,j}$ are their angular positions, 
$w_{i,j}$ is the weight, and
the summation runs over all the pairs of galaxies which reside in a separation bin; $\theta \in |{\bf n}_i-{\bf n}_j|$.
We adopt 24 separation bins logarithmically spaced in the range 
of $\theta=[0.276,332.9]$~arcmin. 
The maximum separation, $\theta=332.9$~arcmin, and the bin width are the same as that of Li+HSCY3.

In this paper, we use the measured $\xi_{\pm}$ down to the smallest scale, $\theta=0.28$~arcmin. 
As shown in Fig.~7 of \citet{HSC3_catalog_Li2021}, the majority of the source galaxies have $R_2<0.9$ for the  
\reGauss{} resolution factor, which is defined for each galaxy using the trace of the second moments 
of the PSF ($T_{\rm PSF}$) and those of the observed galaxy image ($T_{\rm gal}$): $R_2\equiv 
1-T_{\rm PSF}/T_{\rm gal}$. The fact of $R_2<0.9$ means that the HSC galaxies have $T_{\rm gal}\lesssim 
10T_{\rm PSF}$. Since the HSC $i$-band data has a typical seeing size of $0^{\prime\prime}\!\!.6$
\citep{HSC3_catalog_Li2021}, the source galaxies used in the cosmic shear analysis have a size of $r_{\rm gal}\lesssim \sqrt{10}\times r_{\rm PSF}\simeq 
1^{\prime\prime}\!\!.9$. Thus the galaxy size is much smaller than the smallest separation of 
$0'\!.276\simeq 17^{\prime\prime}$ in the 2PCFs.  We do not expect any major systematic errors from using the two galaxies separated by the smallest separation. We also note that 
we use very conservative masks around bright stars \citep{HSC3_catalog_Li2021}, so two galaxies around bright stars, which are rare populations anyway, would not be a major systematics effect. 
We will adopt several choices of small-scale cuts for $\xi_{\pm}$ to study how the results change with the different scale cuts.

\subsection{Covariance}
\label{ssec:covariance}

We estimate a covariance matrix of the measured 2PCFs, $\xi_{\pm}$, using the 1,404 HSC mock shear catalogs constructed using the same method in \citet{HSC1_mock_Shirasaki2019}. The different realizations are constructed using different realizations of cosmic shear, galaxy intrinsic shape, and measurement noise from image noise. We measure the 2PCFs from all 1,404 realizations of the mock catalogs in the same way as the measurement, and then calculate the covariance matrix from these 1,404 measurements. 

Since the cosmic shear signals in the mock catalogs are obtained from a large number of full-sky ray-tracing simulations for the WMAP9 cosmology which takes into account nonlinear structure formation \citep{raytracingTakahashi2017}, the derived cosmic variance includes both Gaussian and non-Gaussian contributions. 
In addition, the galaxy positions and survey geometry in the mock catalogs mimic those of the real HSC data; therefore the derived covariance includes super-survey covariance \citep{2013PhRvD..87l3504T,HSC1_mock_Shirasaki2019}. Moreover, the random shape noise and measurement error are generated using the real shape catalog. We find that the shape noise covariance gives a dominant contribution to the covariance in the small separation bins, while the sample variance dominates the covariance in large separation bins \citep[see Figure~4 in][]{HSC1_mock_Shirasaki2019}.

We find that the average 2PCFs from the mock catalogs are lower than the theory prediction for the reference 
$\Lambda$CDM cosmology and the ratio is approximately constant over the range of angular scales in the fiducial scale cuts (about 0.81) \citep[see Figure~3 in][]{HSC1_mock_Shirasaki2019}, we multiply the 2PCFs from each realization of the mocks by $1/0.81$ in each separation bin (and then calculate the covariance matrix). 

The dimension of the covariance matrix is up to $410\times 410$, where 
we use 180~elements for $\xi_+$ because of
18~separation bins for the smallest scale cut and 10 correlation functions for the four source redshift 
tomography, while 
we use $230$~elements for $\xi_-$ because of 23~separations bins. 
We include the cross covariance between $\xi_+$ and $\xi_-$.
To estimate the inverse of the covariance matrix, which is needed for the likelihood analysis, 
we take into account the Hartlap factor \citep{covariance_Hartlap2007}.

\section{Parameter Estimation Method}
\label{sec:method}

In this section, we introduce our parameter estimation method including the scale cuts and model choices.

\subsection{Scale cuts}
\label{subsec:scale-cuts}

The measured 2PCFs generally include contributions from both curl-free gradient ($E$-mode) and curl component 
($B$-mode). Since the scalar gravitational potential of large-scale structure causes only the $E$-mode in the weak lensing regime, the $B$-mode 2PCFs, reconstructed from the measured $\xi_{\pm}$, can be used as a test of systematic errors. Section~IIIC of Li+HSCY3 gave the reconstructed $B$-mode 2PCF. Based on the result, we will not use $\xi_+$ at $\theta\ge 56.6$~arcmin and $\xi_-$ at $\theta\ge 248$~arcmin, respectively, for the following results, as done in Li+HSCY3.

For the small scales, the $B$-modes are not significant and the scale cuts can be determined to reduce the modeling uncertainties of baryonic effects as done in Li+HSCY3.
As seen in Fig.~\ref{fig:xipm_k-modes}, 
the 2PCFs in smaller separation bins
are sensitive to the higher $k$ modes.
As our nonlinear matter power spectrum emulator, \textsc{DarkEmulator2} achieves a sub-percent accuracy down to $k \simeq 100~h{\rm Mpc}^{-1}$ (see the following section), we can use
the $\xi_{\pm}$ information at the smallest scales. 
To study how the results change with varying the small scale cuts, we consider the following 
four choices of the scale cuts: 
\begin{enumerate}
    \item $\boldsymbol{\theta}_{+{\rm \textbf{min}}}$\textbf{:7'\!.1~(Y3cut)}~:\\
        $\theta_+ \in [7.1, 56.6]$ and $\theta_- \in [31.2, 247.8]$
    \item {$\boldsymbol{\theta}_{+{\rm \textbf{min}}}$\textbf{:2'\!.9}~:}
        $\theta_+ \in [2.94, 56.6]$ and $\theta_- \in [12.9, 247.8]$
    \item $\boldsymbol{\theta}_{+{\rm \textbf{min}}}$\textbf{:1'\!.2}~:
        $\theta_+ \in [1.21, 56.6]$ and $\theta_- \in [3.95, 247.8]$
    \item $\boldsymbol{\theta}_{+{\rm \textbf{min}}}$\textbf{:0'\!.28}~:
        $\theta_+ \in [0.28, 56.6]$ and $\theta_- \in [0.28, 247.8]$
\end{enumerate}
where the numbers in brackets for $\theta_+$ and $\theta_-$ are the ranges of separation bins for 
$\xi_+$ and $\xi_-$, respectively, used in the parameter inference. 
Note that ``Y3cut" is the same as the fiducial scale cuts used in 
Li+HSCY3.
In the following we call ``$\theta_{+{\rm min}}:0'\!.28$'' and so on, to refer the 
respective scale cuts. 
For the cases (1)--(3) of the scale cuts, we adopt a value of $\theta_{-{\rm min}}$ about 4 times larger than 
$\theta_{+{\rm min}}$ for the small-scale cuts for $\xi_-$ and $\xi_+$, based on 
Eq.~(\ref{eq:theta-4theta+}), so that both the two 2PCFs of $\xi_{+/-}$ are sensitive to the similar range of $k$ in $P_{\rm m}(k)$.
For the case (4), we use all available scales in the parameter inference.

\subsection{Model}
\label{ssec:model}

\subsubsection{\textsc{DarkEmulator2}}
\label{sec:dark_emulator}
\begin{table}
\centering
\caption{The
parameter range covered by DQ2 simulations. 
The first five lines depict the varied parameters accepted by the DE2 software as independent inputs,
defining a model within the flat-geometry $\Lambda$CDM framework. 
$A_s$ and $n_s$
are the amplitude and spectral tilt parameters of the primordial curvature power spectrum at 
pivot scale $k_{\rm pivot}=0.05~{\rm Mpc}^{-1}$, $h$ is the Hubble constant parameter, and $\omega_{\rm b}(\equiv \Omega_{\rm b}h^2)$ is the physical density parameter of baryon. 
The physical density parameter of CDM is given as $\omega_{\rm c}(\equiv \Omega_{\rm c}h^2)=\Omega_{\rm m}h^2-\omega_b-\omega_\nu$,
where $\omega_\nu$ is the physical density parameter of massive neutrinos.
The density parameter of the matter ($\Omega_{\rm m}$), the dark energy ($\Omega_{\rm de}$), and curvature ($\Omega_{\rm k}$) satisfy $\Omega_{\rm m} +\Omega_{\rm de}  + \Omega_{\rm k} = 1$.
Note that we apply further cuts based on the parameter combination, $S_8$, in the $\Omega_\mathrm{m}$-$\sigma_8$ plane, resulting in a subspace with a ``banana shape'' rather than a rectangle.
The subsequent four parameters are also varied in DQ2 but are held constant in this paper. The four parameters in the next group are automatically determined once the first nine are specified. Additionally, the redshift and wavenumber ranges are presented at the bottom.
}
\label{tab:cosmological_parameters_supportingrange}
  \begin{tabular}{ccc}
   \hline \hline
   Parameter & range & constant value \\
   \hline
   $\Omega_\mathrm{m}$ & [0.05, 0.62] & -- \\
   $\omega_\mathrm{c}$ & [0.01, 0.3] & -- \\
   $\sigma_8$ & [0.47, 1.23] &  -- \\
   $n_\mathrm{s}$ & [0.916, 1.012] & -- \\
   $h$ & [0.5, 0.9] & -- \\
   $M_\nu \, [\mathrm{eV}]$ & [0.0, 0.5] & 0.06 \\
   $w_0$ & $[-1.5, -0.5]$ & $-1$ \\
   $w_a$ & $[-0.5, 0.5]$ & 0 \\
   $\Omega_\mathrm{k}$ & $[-0.1, 0.1]$ & 0 \\
   \hline 
   $\omega_\mathrm{b}$ & [0.015, 0.03] & -- \\
   $\Omega_\mathrm{de}$ & [0.4125, 0.9385] & -- \\
   $A_\mathrm{s}$ & [$2.833 \times 10^{-10}$, $3.157\times 10^{-8}$] & -- \\
   $S_8$ & [0.6, 0.95] & -- \\ 
  \hline
$z$ & $[0, 3]$ & -- \\
$k \, [h \mathrm{Mpc}^{-1}]$ & $[0.001, 100]$ & -- \\
\hline\hline
  \end{tabular}
\end{table}

\begin{table}
\begin{center}
\caption{{Basic simulation parameters for the DQ2 simulations: the number of simulation particles ($N_\mathrm{p}$), comoving box size in $h^{-1}\mathrm{Mpc}$ ($L_\mathrm{box}$), number of cosmological models sampled ($N_\mathrm{model}$) and the particle Nyquist wavenumber in $h\,\mathrm{Mpc}^{-1}$ ($k_\mathrm{ny} = \pi N_\mathrm{p}^{1/3}/L_\mathrm{box}$).
\label{tab:DQ2_spec}}}
\begin{tabular}{cccc} \hline\hline
$N_\mathrm{p}$ & $L_\mathrm{box}$ & $N_\mathrm{model}$
& $k_\mathrm{ny}$\\ \hline
$1024^3$ & $1024$ & $1000$ & $3.14$ \\
$2048^3$ & $1024$ & $50$ & $6.28$ \\
$3000^3$ & $1000$ & $20$ & $9.42$ \\
\hline\hline
\end{tabular}
\end{center}
\end{table}

In this section, we briefly describe our nonlinear matter power spectrum emulator, \textsc{DarkEmulator2}. A more detailed description of this software is presented in a separate paper ({Tanaka, Nishimichi \& Kobayashi in prep.}). The emulator is based on a new simulation campaign, \textsc{Dark Quest~2} (DQ2), the successor of a previous project, \textsc{Dark Quest~1} \cite{darkemu_Nishimichi2019}. The main focus of the new simulations is to encompass a broad parameter space in a nine-parameter $w_0w_a o \nu$CDM model, with a time-varying equation-of-state parameter for dark energy ($w_0, w_a$), a nonzero spatial curvature ($\Omega_{\rm k}$), and the sum of the three neutrino masses ($M_{\nu}$). Focusing primarily on the typical constraints from cosmic shear measurements and potentially broadened $\Omega_\mathrm{m}$-$\sigma_8$ parameter subspace in extended models compared to the constraints for the vanilla-$\Lambda$CDM, the DQ2 simulations employ broader ranges for these parameters in a ``banana-shaped'' region to fully cover the HSC-Y1 cosmic-shear constraints, roughly up to the $3$-$\sigma$ range~\cite{cosmicShear_HSC1_Hamana2019}. We show in Table~\ref{tab:cosmological_parameters_supportingrange} the range of cosmological parameters adopted in DQ2.

For efficient emulator construction, a mixed learning simulation dataset with three different mass/spatial resolutions 
is used (see Table~\ref{tab:DQ2_spec} for details). The lowest-resolution simulations are performed at $1,000$ cosmological models to capture the dependence in the high-dimensional input-parameter space, while $50$ ($20$) cosmologies are used for the middle (high) resolution simulations to further capture the resolution-dependent features, particularly enhancing accuracy on small scales. The neural-network based emulator learns not only the dependence of the matter power spectrum on cosmological parameters but also its resolution dependence. In turn, it accepts an input variable that specifies the resolution. We use the predictions for the high-resolution setup in this paper. 

The simulations are performed with the \textsc{GINKAKU} code ({Nishimichi, Tanaka, \& Yoshikawa in prep.}), which is a Tree-Particle Mesh code developed based on the Framework for Developing Particle Simulator (\textsc{FDPS}) library~\cite{2016PASJ...68...54I,2018PASJ...70...70N}. The accuracy parameters in the simulation code are tuned to reproduce the simulations by \textsc{Gadget2}~\cite{gadget1,gadget2} with the parameters adopted in Refs.~\cite{halofitT2012,darkemu_Nishimichi2019}. The matter power spectra are denoised on large scales based on a treatment updated from the one described in \cite{darkemu_Nishimichi2019} using the 
propagator method based on perturbation theory \cite{crocce:2006uq}. The final accuracy of the mixed-resolution model is evaluated as $\sim 0.4\%$ averaged over redshifts ($0\leq z \leq 3$) and wavenumbers ($0.001 < k/[h\mathrm{Mpc}^{-1}] < 100$) using test simulations, which are not used either in training or in validation of the model.

To compute the cosmic-shear 2PCFs or power spectra, we need a model for the matter power spectrum over a wide range of wavenumbers. However, power spectra on small scales well above the particle Nyquist wavenumber are still contaminated by noise stemming from the discreteness of particles even for the high-resolution setup. This contamination primarily depends on the resolution of the simulations and the redshift, with a weak dependence on the cosmological parameters, affecting both the simulation data and consequently the emulator predictions, as the latter are built on the former. In order to mitigate this effect when evaluating cosmic shear signals, we identify the wavenumber at which the power spectrum exceeds ten times the Poisson noise level. Beyond this wavenumber, we extrapolate the spectrum assuming a single power law, rather than relying on the direct output of the neural network. The error introduced by this treatment can be estimated by running the code with different resolution parameters, as it tends to diminish as resolution increases. We have verified that our conclusions remain robust regardless of the choice of the resolution parameter or against the implementation details of this treatment.
As shown in Fig.~\ref{fig:xipm_k-modes}, the cosmic shear 2PCFs on the angular scales we use in this paper are sensitive to the matter power spectrum at $k\lesssim 20~h{\rm Mpc}^{-1}$, so we believe that the above extrapolation has little effect on our results.

\subsubsection{Model templates}

We use the following model template of cosmic shear 2PCFs for source galaxies in the $i$- and $j$-th redshift bins, for each input model: 
\begin{widetext}
\begin{align}
\xi^{({\rm t})ij}_{\pm}(\theta|\mathbb{C},\Delta z_{i},\Delta m_i,\alpha_{\rm psf}^{(2)},\alpha_{\rm psf}^{(4)},\beta_{\rm psf}^{(2)},\beta_{\rm psf}^{(4)})
=(1+\Delta m_i)(1+\Delta m_j) \xi^{ij}_{\pm}(\theta|\mathbb{C},\Delta z_{i})
+\xi_{\rm psf}(\theta|\alpha_{\rm psf}^{(2)}, \alpha_{\rm psf}^{(4)},\beta_{\rm psf}^{(2)}, \beta_{\rm psf}^{(4)}),
\label{eq:model_template}
\end{align}
\end{widetext}
where $\mathbb{C}$ is a set of cosmological parameters that specify an input flat $\Lambda$CDM model, 
the first term on the r.h.s. is for the cosmic shear signal that is computed using Eqs.~(\ref{eq:model_hankel}) and (\ref{eq:cl_EE}) for the input 
$\Lambda$CDM model, and the second term models the contamination arising from the systematic PSF modeling. 
\citet{2023MNRAS.525.2441Z} quantified 
the PSF systematic errors using the cross-correlations between the galaxy shapes and the shape modeling errors of PSF stars. 
$\alpha_{\rm psf}^{(2)}$ and $\alpha_{\rm psf}^{(4)}$ parametrize the PSF leakage bias by the PSF second- and fouth-order moments, and $\beta_{\rm psf}^{(2)}$ and $\beta_{\rm psf}^{(4)}$ parametrize the PSF modeling error in the second- and fourth-order moments. We
use the same method described in Section~IVE of Li+HSCY3.
The factor $(1+\Delta m_i)(1+\Delta m_j)$ models contamination of the multiplicative shear bias
for each sample of source galaxies in the tomographic bins. 
$\Delta z_i$ ($i=1,2,3$ or 4) is a parameter to 
model residual uncertainties in the mean of source redshifts for each tomographic bin.

We note that for the cosmic shear term in Eq.~(\ref{eq:model_template}), we consider only the $E$-mode contribution, that is, the weak lensing contamination, and, in other words, we ignore the $B$-mode contamination that could arise from other effects such as IA, baryonic effect and nonlinear lensing effects. In this sense, the model template we use is the {\em simplest} one and therefore contains {\em minimal} modeling uncertainty. We will discuss later how adding the other model ingredients such as a model of the baryonic effect changes our results.

\subsubsection{Bayesian inference}
\label{sec:bayes}

\begin{table}
\caption{
Model parameters and priors used in our fiducial cosmological parameter
inference. The label ${\mathcal U}(a,b)$ denotes a uniform flat prior
between $a$ and $b$, and ${\mathcal N}(\mu, \sigma)$ denotes a normal
distribution with mean $\mu$ and width $\sigma$.
Our fiducial analysis uses: five cosmological parameters, eight 
nuisance parameters to model residual error in the mean redshift and 
multiplicative shear bias for each source galaxy population of 
the four tomographic bins, and four parameters to model residual PSF 
modeling errors in the cosmic shear 2PCF (see text for details). We have 17 parameters in total. 
The number shown in bold font denotes the number set by the support range of 
\textsc{DarkEmulator2} emulator, which is different from that used in 
Li+HSCY3.
}
\label{tab:parameters}
\setlength{\tabcolsep}{20pt}
\begin{center}
\begin{tabular}{ll}  \hline\hline
Parameter & Prior \\ \hline
\multicolumn{2}{l}{\hspace{-1em}\bf Cosmological parameters}
\\
$\Omega_\mathrm{m}$                 & ${\cal U}(0.1, \bf{0.62})$\\
$A_\mathrm{s} \,(\times 10^{-9})$   & ${\cal U}(0.5, 10)$\\
$n_\mathrm{s}$                      & ${\cal U}(\bf{0.92}, \bf{1.01})$\\
$h$                               & ${\cal U}(0.62, 0.80)$\\
$\omega_\mathrm{b}$                 & ${\cal U}(0.02, 0.025)$\\
\hline
\multicolumn{2}{l}{\hspace{-1em}\bf Photo-$z$ systematics (see Section~\ref{sssec:photoz})}\\
$\Delta z_{1}$                    & ${\cal N}(0, 0.024)$ \\
$\Delta z_{2}$                    & ${\cal N}(0, 0.022)$ \\
$\Delta z_{3}$                    & {${\cal N}(-0.117, 0.052)$} \\
$\Delta z_{4}$                    & {${\cal N}(-0.199, 0.088)$} \\
\multicolumn{2}{l}{\hspace{-1em}\bf Shear calibration biases (see Eq.~\ref{eq:model_template})}\\
$\Delta m_{1}$                    & ${\cal N}(0.0,0.01)$ \\
$\Delta m_{2}$                    & ${\cal N}(0.0,0.01)$ \\
$\Delta m_{3}$                    & ${\cal N}(0.0,0.01)$ \\
$\Delta m_{4}$                    & ${\cal N}(0.0,0.01)$ \\
\multicolumn{2}{l}{\hspace{-1em}\bf PSF systematics
(see Eq.~\ref{eq:model_template})}\\
$\alpha'^{(2)}$                   & ${\cal N}(0, 1)$\\
$\beta'^{(2)}$                    & ${\cal N}(0, 1)$\\
$\alpha'^{(4)}$                   & ${\cal N}(0, 1)$\\
$\beta'^{(4)}$                    & ${\cal N}(0, 1)$\\ \hline
\hline
\end{tabular}
\end{center}
\end{table}
We use a Monte Carlo Bayesian analysis to sample the posterior in the 17~dimensional space of 
the cosmological, astrophysical and systematic parameters. 
We adopt a Gaussian likelihood ${\cal L}$:
\begin{align}
-2\ln{\cal L}(\hat{\xi}_{\pm}|{\bf p})=\left(\hat{\xi}^{ij}_{\pm}-\xi_{\pm}^{({\rm t})ij}({\bf p})\right)^T
{\bf C}^{-1}
\left(\hat{\xi}^{ij}_{\pm}-\xi^{({\rm t})ij}_{\pm }({\bf p})
\right),
\end{align}
where ${\bf p}$ denotes the vector of parameters and ${\bf C}^{-1}$ is the inverse of the covariance matrix. 
The 2PCFs have up to 410 dimensions for the scale cut of the smallest scales
(18 separation bins for each of the 10 correlation functions of $\xi_+$,
and 23 separation bias for each of $\xi_-$'s, respectively).

With Bayesian inference, we can estimate a posterior probability distribution, denoted
as ${\cal P}({\bf p}|\hat{\xi}_{\pm})$, for the parameters ${\bf p}$, given the data vector $\hat{\xi}_{\pm}$:
\begin{align}
{\cal P}({\bf p}|\hat{\xi}_{\pm})\propto{\cal L}(\hat{\xi}_{\pm}|{\bf p})\Pi({\bf p}),
\label{eq:bayesian}
\end{align}
where $\Pi({\bf p})$ is the prior distribution of parameters. 

As described above, we include the nuisance parameters $\Delta z_i$ ($i=1,2,3$ or 4) to 
account for residual uncertainties in the mean of source redshifts.
For the first and second tomographic bins we adopt the informative Gaussian prior with width that was 
estimated from photo-$z$ information for each source population as given in 
\citet{HSC3_photoz_Rau2022}. These Gaussian priors are the same as those used in Li+HSCY3.
For the third and fourth tomographic bins, Li+HSCY3 found an indication of residual systematic errors in the mean of source redshifts. More precisely, they found that, by adopting the uniform prior on $\Delta z_3$ and $\Delta z_4$, the posteriors are not consistent with zero, indicating that the true redshift distributions of the last two tomographic bins are higher by $\Delta z\sim 0.1$ than estimated by the photo-$z$ method of \cite{HSC3_photoz_Rau2022}. This finding was 
done in a self-calibration manner in the sense that the relative amplitudes and shapes of the cosmic shear 2PCFs, relative to those for the lower redshift bins, enabled to calibrate the mean redshifts, for flat $\Lambda$CDM cosmologies. 
Since the main purpose of this paper is to study a possible signature of the baryonic effect in the cosmic shear 2PCFs, we employ the informative Gaussian priors on $\Delta z_3$ and $\Delta z_4$ with mean and width that are obtained by the same scale cut (``$\theta_{+{\rm min}}:7'\!.1$~arcmin": Y3cut) as in Li+HSCY3,
as we will below explain in detail. Note that, although our fiducial analysis setup is not exactly the same as that of Li+HSCY3,
the posteriors of $\Delta z_3$ and $\Delta z_4$ are very similar to those of Li+HSCY3.
We will also discuss how adopting the flat uniform priors on $\Delta z_3$
and $\Delta z_4$ changes our results when using the smaller scale cuts than in Li+HSCY3.

We also include a nuisance parameter, $\Delta m_i$ ($i=1,2,3$ or 4), to model the uncertainties from the multiplicative shear bias residuals, for the $i$-th source galaxy population. The parameters thus model the 
redshift-dependent multiplicative bias residual. We employ the Gaussian prior of $\Delta m_i$ with zero mean 
and a standard deviation of 0.01, which is motivated by the calibration of the HSC-Y3 shear catalog based on image simulations \citep{HSC3_catalog_Li2021} since it is confirmed that the multiplicative bias residual is controlled below the 1\% level. The treatment of $\Delta m_i$ is the same as in 
Li+HSCY3.

Our analysis uses a set of parameters and priors summarized in Table~\ref{tab:parameters}.
The parameters include five cosmological parameters, denoted by $\mathbb{C}=\left\{\Omega_{\rm m}, 
A_s, n_s, h, \omega_{\rm b}\right\}$,
for flat $\Lambda$CDM cosmologies {(see the caption of Table.\ref{tab:cosmological_parameters_supportingrange} for definition)}. As in Li+HSCY3, 
we adopted the fixed total neutrino mass, $M_\nu=0.06~{\rm eV}$. 

Using the method given in Section~IV~F of Li+HSCY3,
we performed the transformation of the original PSF error parameters 
to obtain the new set of parameters in that the mean of each new parameter is shifted to zero and  
the covariance matrix of the new parameters becomes diagonal. We then employ a Gaussian prior for each of the new uncorrelated parameters, $\{\alpha^{\prime(2)},\beta^{\prime(2)},\alpha^{\prime(4)},\beta^{\prime(4)}\}$. 
We normalize the distribution width of each parameter by the measured width in \citet{2023MNRAS.525.2441Z}, so we employ a Gaussian prior with 
zero mean and unit width: ${\cal N}(0,1)$.

We use \textsc{PolyChord} to analyze the data and noiseless mocks, as done in Li+HSCY3,
to obtain the posterior distribution of parameters. We also use 
\textsc{MultiNest} to analyze noisy mocks (in Section~\ref{subsec:goodness of fit}) because 
\textsc{MultiNest} is about five times faster than 
\textsc{PolyChord}. Note that the marginalized posterior widths such as those for $\sigma_8$ and $\Omega_{\rm m}$ estimated by 
\textsc{MultiNest} are about 10\% smaller than those from 
\textsc{PolyChord}. However, since we are only interested in point estimates such as the MAP and 1D marginalized mode for the noisy mock analysis, the effects of this underestimation of posterior widths would be marginal. We also note that all the posteriors shown in this paper are obtained using \textsc{PolyChord}.

Throughout this paper, we report the 1D marginalized {\em mode} and its asymmetric 68\% credible intervals,
together with the MAP estimated as the maximum of the posterior in the chain
\citep[see Eq.~1 in][]{HSC1_2x2pt_Miyatake2022}. For the 2D marginalized posterior, 
we report the mode and the 68\% and 95\% credible intervals. 
We use \textsc{GetDist} \citep{GetDist2019} for the plotting.

\section{Results}
\label{sec:results}

\subsection{Consistency test with HSCY3 cosmology result}
\label{sec:consistency_with_hscy3}

\begin{figure}
    \includegraphics[width=\columnwidth]{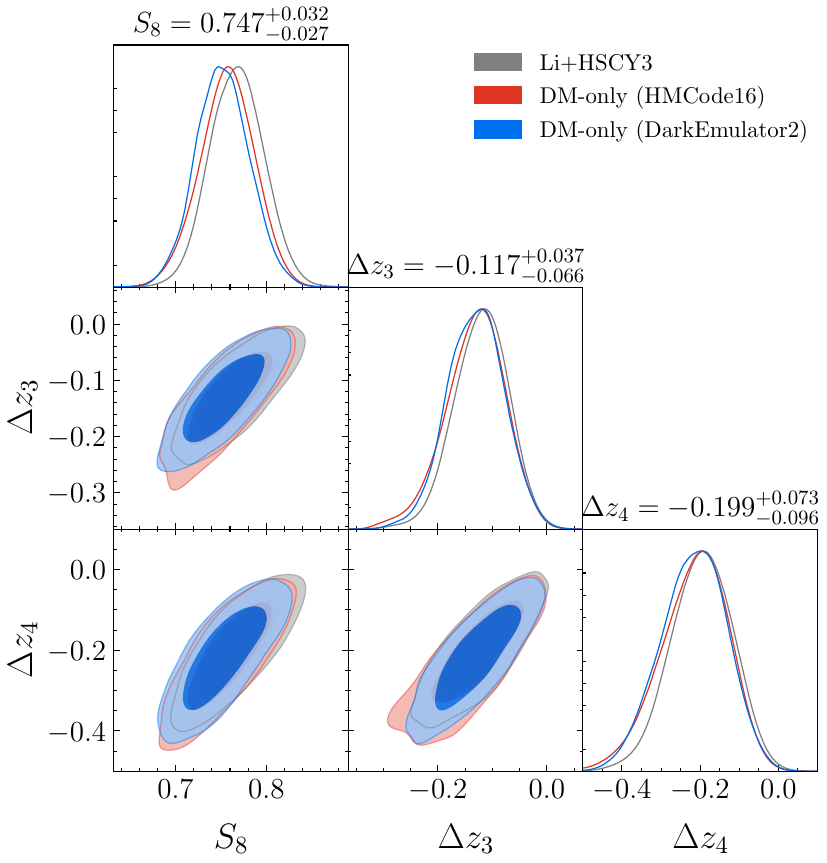}
    \caption{Consistency test of our analysis method with the published HSC-Y3 cosmology result of Li+HSCY3, using the fiducial scale cut of Li+HSCY3, 
    ``$\theta_{+{\rm min}}=7'\!.1$'' given in Section~\ref{subsec:scale-cuts}. 
    The darker and lighter shaded regions denote the marginalized 68\% and 95\% credible intervals. 
    For all the results, we use flat priors on $\Delta z_3$ and 
    $\Delta z_4$, which model residual uncertainties in the mean redshift 
    of the third and fourth tomographic bins. 
    The gray-color shaded region or line 
    reproduces the 
   posteriors for the parameters of ($S_8, \Delta z_3, \Delta z_4$). 
        Li+HSCY3 used the \textsc{HMCode16}~\citep{halofit_mead16} to model the nonlinear matter power spectrum, which includes the parameters to model the baryonic effect. 
    The blue posteriors denote the results based on our analysis method, which uses 
    the \textsc{DarkEmulator2} 
    emulator to model the DM-only matter power spectrum (i.e. ignoring the baryonic effect). For comparison, the red-color posteriors 
    show the results obtained by switching off the baryonic effect in the \textsc{HMCode16} model, 
    which corresponds to the DM-only model in the Li+HSCY3 analysis setup.
    For all the results we include the TATT IA model in the model template. 
    The number and the error bars above each panel 
    of the 1D posterior denote the central value and 
    the 68\% credible interval for the parameter,
    for the ``DarkEmulator2'' result.
    }
    \label{fig:2D_LiHSCY3}
\end{figure}

Before going to the main results, we first show how different the cosmological parameters obtained 
by our analysis method are in comparison with those of 
Li+HSCY3, when using a similar analysis setup to that in Li+HSCY3.
Recall that Li+HCY3 used the \textsc{HMCode16}~\citep{halofit_mead16} including the parameter to model the baryonic effect on the matter power spectrum.
They used the tidal alignment and tidal torque (TATT) model of the IA effect~\citep{tatt_blazek17}, which is given by   
five parameters (see Section~\ref{subsec:IA} and also Section~IVB of Li+HSCY3).
In addition, Li+HSCY3 adopted uninformative flat priors on the residual uncertainties in the mean redshifts of the third and fourth tomographic bins, denoted as $\Delta z_3$ and $\Delta z_4$ (see Eq.~\ref{eq:dz_def}).
To compare with Li+HSCY3, 
here we adopt the same flat priors on $\Delta z_3$ and $\Delta z_4$, the same TATT IA model,  
and the same scale cuts (``$\theta_{+{\rm min}}:7'\!.1$'' in Section~\ref{subsec:scale-cuts})
as used in Li+HSCY3, but use the DM-only model, \textsc{DarkEmulator2}, of the nonlinear matter power spectrum to compute the cosmic shear 
2PCFs for an input $\Lambda$CDM model.

Fig.~\ref{fig:2D_LiHSCY3} shows the results for the consistency test with Li+HSCY3. 
The gray-color posterior distributions show the results of Li+HSCY3, i.e. the HSC-Y3 results. 
The blue-color posteriors show the results from our analysis method. For comparison, the red-color 
posteriors show the results when ignoring the baryonic effect, i.e. using the fixed value of $A_b=3.13$ in the \textsc{HMCode16}, which corresponds to the DM-only model of the matter power spectrum. The posteriors are almost unchanged, meaning that the baryonic effect is unlikely to be significant over the range of scales defined by the fiducial scale cuts, 
as also carefully studied in Li+HSCY3. 
In contrast to these, the blue-color posteriors show the results of our method. A subtle change in the posterior 
of $S_8$, compared to the red-color posterior, is due to the difference in the nonlinear matter power spectrum that is modeled by the \textsc{DarkEmulator2} model in our method compared to the \textsc{HMCode16}.
On the other hand, the posteriors of $\Delta z_3$ and $\Delta z_4$ are almost unchanged. 
To be more precise, we find $\Delta z_3 = -0.117^{+0.037}_{-0.066}$ and $\Delta z_4 = -0.199^{+0.073}_{-0.096}$ for our method,
which is very close to $\Delta z_3=-0.115^{+0.052}_{-0.058}$ and $\Delta z_4=-0.192^{+0.088}_{-0.088}$ of Li+HSCY3. For the following results using our fiducial analysis setup, 
we take the central value and the 68\% credible interval in the above for 
 the mean and width of Gaussian 
priors on $\Delta z_3$ and $\Delta z_4$ as given in Table~\ref{tab:parameters}.

\subsection{Cosmological constraints with different scale cuts}
\label{sec:cosmology_constraints}

\begin{figure}
    \includegraphics[width=\columnwidth]{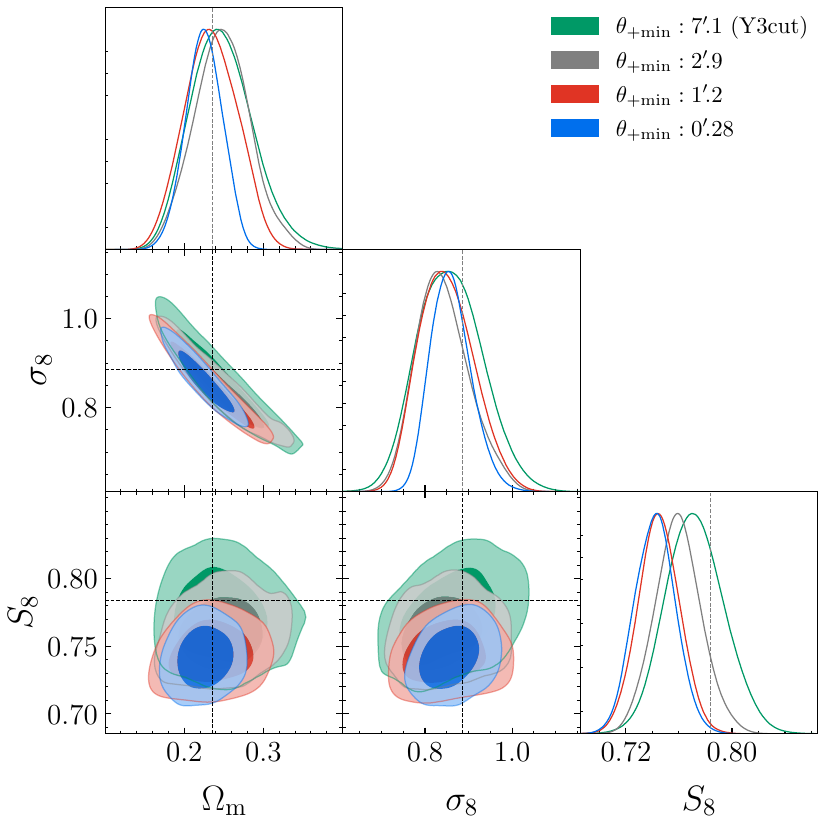}
    \caption{The 1D and 2D posteriors obtained from the parameter inference of the ``noiseless'' mock data vector simulated using the \textsc{HMCode20} at the MAP model with  
    $\Theta_{\rm AGN}=8.0$ (yellow-dashed line in Fig.~\ref{fig:xipm_DM_mead20}). 
    For the analysis setup,
    we use the fiducial setup in Table~\ref{tab:parameters}, i.e. using the DM-only
    \textsc{DarkEmulator2}  model in the model template. The vertical and horizontal lines in each panel show the input value of 
    the parameter used for the noiseless mock data. 
    The central value of $S_8$ shows a systematic shift when using the smaller scale cuts as denoted by the legend.  
    }
    \label{fig:2D_TAGN8_noiselessmock}
\end{figure}

\begin{figure*}
    \includegraphics[width=1.8\columnwidth]{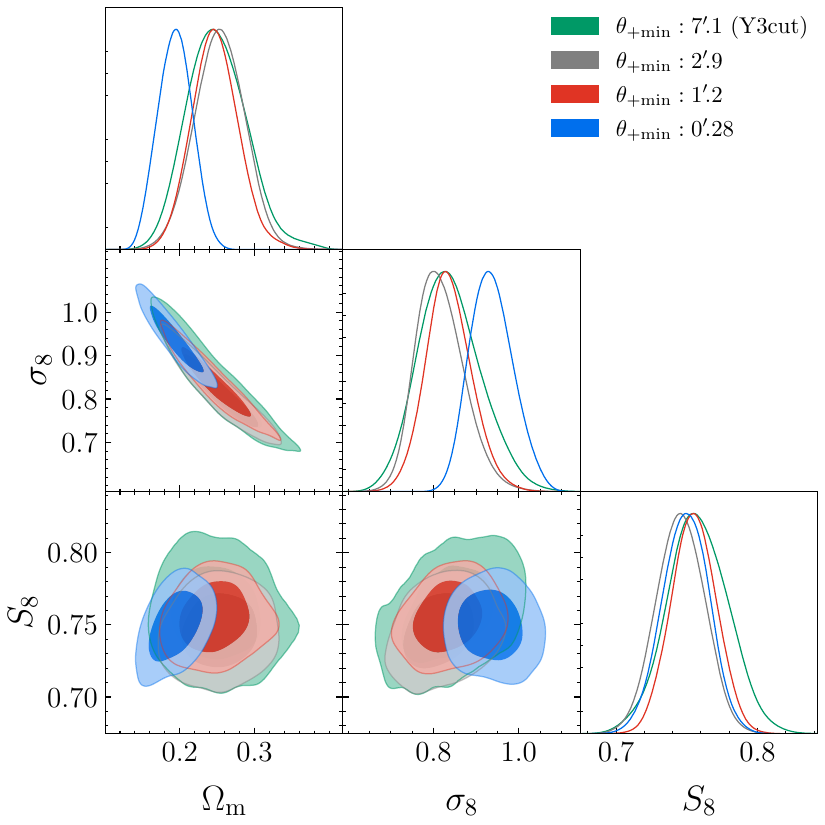}
    \caption{The marginalized 1D and 2D posteriors of the parameters $(\Omega_{\rm m},\sigma_8,S_8)$, obtained 
    by applying our fiducial analysis to the HSC-Y3 cosmic shear 2PCFs: 
    we use the 
    \textsc{DarkEmulator2} emulator to model the matter power spectrum ignoring 
    the baryonic effect (i.e. DM-only model prediction) and use a set of parameters and the priors given in Table~\ref{tab:parameters}.
    The different color posteriors show the results using the different scale cuts denoted in the legend 
    (see Section~\ref{subsec:scale-cuts}). Note that we do not include the IA effect in the model template for all the results. 
    }
    \label{fig:2D}
\end{figure*}
\begin{figure*}
    \includegraphics[width=1.8\columnwidth]{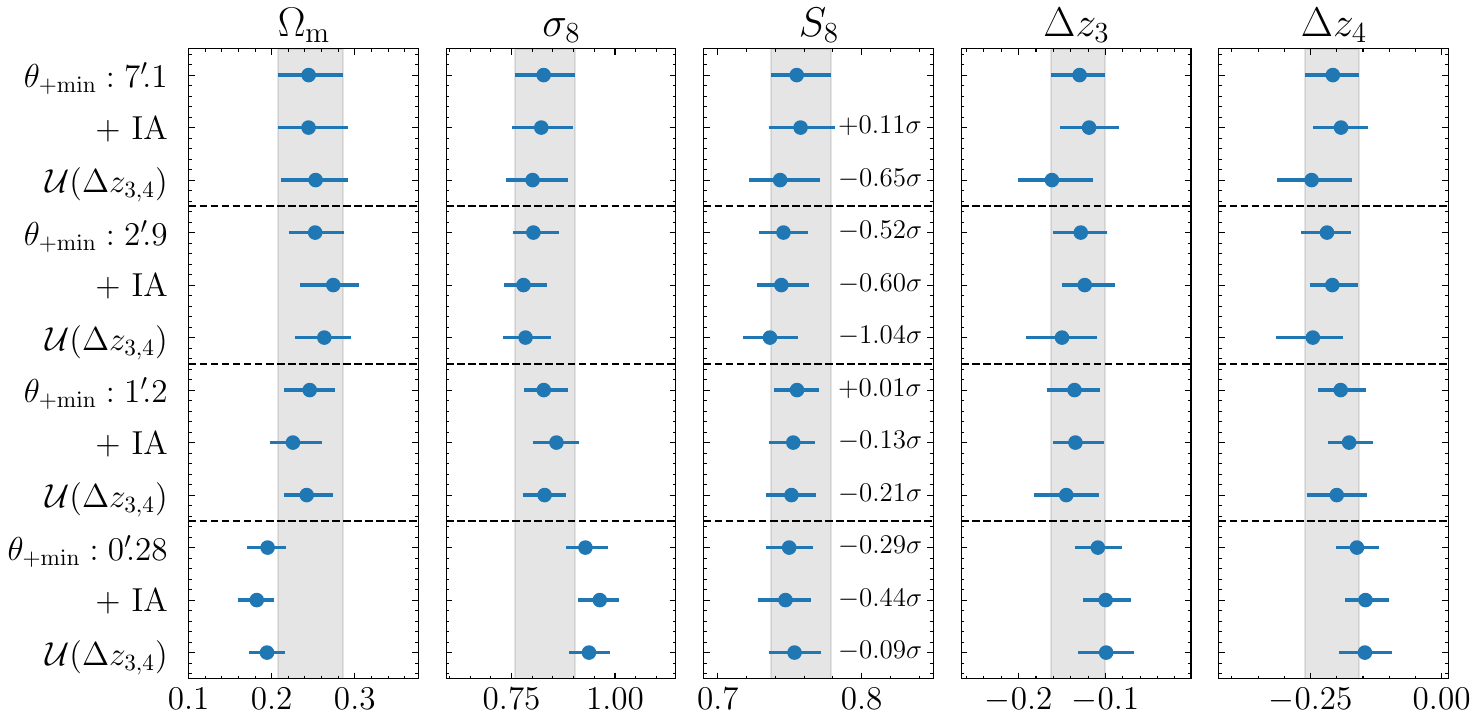}
    \caption{The central
     (mode) value and 68\% credible interval of the 
    parameters, $\Omega_{\rm m}$, $\sigma_8$, $S_8$, $\Delta z_3$ and $\Delta z_4$ for the different scale cuts as in the previous figure.
    The shaded region for each parameter denotes the result when using the same scale cut 
    (``$\theta_{+{\rm min}}=7\!.1$'') as used in Li+HSCY3.
    The row labeled as ``$\theta_{+{\rm min}}: X'\!.X$" shows the results obtained using the fiducial analysis setup 
    (Table~\ref{tab:parameters}). The results in the rows labeled as ``+IA'' and 
    ``${\cal U}(\Delta z_{3,4})$'' are discussed in detail in Section~\ref{sec:discussion}.
    The  ``+IA'' row shows the results obtained when 
    including the tidal alignment and tidal torque (TATT) model of the IA effect in the model template, which is given by five parameters (see Table~\ref{tab:IA_parameters}).
    The  ``${\cal U}(\Delta z_{3,4})$'' row shows the results when using the uninformative flat priors on $\Delta z_3$ and $\Delta z_4$, instead of the Gaussian priors in Table~\ref{tab:parameters}, which model residual systematic errors in the mean of source redshifts in the third and fourth redshift bins. Note that, in this case, we have not include the IA effect in the model template. 
    The number above each result of $S_8$ for each analysis setup denotes the shift in the central value of $S_8$ with respect to  the central value and the $1\sigma$ width 
    of the fiducial analysis with 
    the ``$\theta_{+{\rm min}}:7'\!.1$'' scale cut.
    }
    \label{fig:1D_IA}
\end{figure*}

We study how our method using the model of the DM-only matter power spectrum changes the cosmological parameters, compared to the fiducial cut (``$\theta_{+{\rm min}}: 7'\!.1$''), when using the different scale cuts.
If the HSC-Y3 data is perfectly described by the DM-only model down to the smallest scale,
the $S_8$ values obtained from the fitting of  
the DM-only model for the 
different scale cuts should be consistent with each other to within the statistical errors,  
hence we expect no significant shift in the best-fit $S_8$ value. 
On the other hand, if the data is affected by the suppression in the matter power spectrum at small scales due to the baryonic effect, which cannot be described by the DM-only model, 
we expect to find lower $S_8$ values when using the smaller scale cuts, 
because the suppression is stronger at smaller scales as shown in Fig.~\ref{fig:xipm}.

Fig.~\ref{fig:2D_TAGN8_noiselessmock} shows possible results expected when the HSC-Y3 data is affected by the baryonic effects, as we described above. Here we use the ``noiseless'' mock data vector generated using the MAP model that is obtained when 
performing the parameter inference of the HSC-Y3 cosmic shear 2PCFs with the \textsc{HMCode20}
model with the fixed AGN parameter $\Theta_{\rm AGN}=8.0$,
for the smallest scale cut ``$\theta_{+{\rm min}}:0'\!.28$'' 
(see Fig.~\ref{fig:xipm_DM_mead20}). 
Then we analyze the mock data using  
the fiducial analysis setup using the DM-only \textsc{DarkEmulator2} model, for the different scale cuts.
The figure shows
a systematic shift in the central $S_8$ values when using the smaller scale cuts, compared to the $S_8$ value of ``$\theta_{+{\rm min}}:7'\!.1$'' (Y3cut), as expected.
Note that the $S_8$ shift 
for the smallest scale cut of ``$\theta_{+{\rm min}}:0'\!.28$'' is similar to that for 
``$\theta_{+{\rm min}}:1'\!.2$'' (or the $S_8$ values are similar for these two scale cuts), 
because $\xi_-$ has an increased power (up-turn shape) compared to the DM-only model prediction at scales 
$\theta \lesssim1$~arcmin, as shown in Fig.~\ref{fig:xipm}, and the inclusion of these small-scale signal compensates for the shift of $S_8$.

Fig.~\ref{fig:2D} shows the main results of this paper, i.e. 
the posterior distributions of $\Omega_{\rm m}$, $\sigma_8$, and $S_8$ obtained from the HSC-Y3 data analysis when using the different scale cuts in our DM-only analysis method. First, 
the credible intervals shrink as we use the smaller scale cuts, meaning that the small-scale 2PCFs carry the statistical power. The use of the smallest scale cut (``$\theta_{+{\rm min}}:0'\!.28$'') leads to a sizable shift in the central values of $\sigma_8$ and $\Omega_{\rm m}$ 
compared to other scale-cut results. However, the central value of $S_8$, which is the most sensitive parameter of the cosmic shear 2PCFs,  
remains largely unchanged. 
 This means that the HSC-Y3 cosmic shear signal does not indicate a significant signature of the baryonic effect. 

Fig.~\ref{fig:1D_IA} shows the central value and the 68\% credible interval for each parameter for different scale cuts. 
All the parameters do not largely change when using the scale cuts down to ``$\theta_{+{\rm min}}:1'\!.2$'', from those of the HSCY3 fiducial cut. For the smallest scale cut, 
the parameters other than $S_8$ change by about $1\sigma$. However, we note that shifts in $\sigma_8$
and $\Omega_{\rm m}$ are still along the degeneracy direction in the cosmic shear signal as can be found from Fig.~\ref{fig:2D}. 
The baryonic effect, if it exists, would not necessarily cause shifts along the degeneracy direction
corresponding to the similar $S_8$, as shown in Fig.~\ref{fig:2D_TAGN8_noiselessmock}. 

Almost no shift in $S_8$ might indicate an undetectable signature of the baryonic effect or other systematic effects in the HSC-Y3 2PCFs at small scales, compared to the statistical measurement errors.
We will further discuss the indication of 
almost no shift in $S_8$ in the light of the baryonic feedback
in the Section~\ref{subsec:S8shift}.

\subsection{Goodness-of-fit}
\label{subsec:goodness of fit}

\begin{table}
\caption{Estimate of the goodness-of-fit of the MAP model obtained from the different analysis setups. 
The column ``scale cut'' denotes the scale cut used in the analysis. The column ``analysis choice'' denotes which analysis setup is used for the parameter inference: 
``fiducial'' means our fiducial method, i.e. using the DM-only model of matter power spectrum as given in Table~\ref{tab:parameters}, ``+IA'' means that we include the TATT IA effect in the model template, and ``${\cal U}(\Delta z_{3,4})$'' denotes the analysis where the flat priors on 
$\Delta z_3$ and $\Delta z_4$, instead of the Gaussian priors in Table~\ref{tab:parameters}, are used. 
The column  ``$\chi^2_{\rm MAP}$'' denotes the $\chi^2$ value at MAP for each method. The column ``$p$-value'' denotes the $p$-value of the MAP model that is estimated from the 50 noisy mocks (see main text for details).
}
\label{tab:goodness-of-fit}
\setlength{\tabcolsep}{8pt}
\begin{center}
\begin{tabular}{llll}  \hline\hline
scale cut & analysis choice & $\chi^2_{\rm MAP}$  & $p$-value ($\chi^2_{\rm MAP}$) \\ \hline
$\theta_{+\rm{min}}:7'\!.1$ & fiducial & 152.2 & 0.10 \\
 & + IA & 150.7 & 0.08 \\
 & $\mathcal{U}(\Delta z_{3,4})$ & 153.6 & 0.11 \\
 \hline
$\theta_{+\rm{min}}:2'\!.9$ & fiducial & 204.4  & 0.20 \\
 & + IA & 202.5 & 0.19 \\
 & $\mathcal{U}(\Delta z_{3,4})$ & 206.2 & 0.17 \\
 \hline
$\theta_{+\rm{min}}:1'\!.2$ & fiducial & 304.5  & 0.03 \\
 & + IA & 300.6 & 0.04 \\
 & $\mathcal{U}(\Delta z_{3,4})$ & 302.1 & 0.03 \\
 \hline
$\theta_{+\rm{min}}:0'\!.28$ & fiducial & 460.3 & 0.02 \\
 & + IA & 451.3 & 0.05  \\
 & $\mathcal{U}(\Delta z_{3,4})$ & 460.7 & 0.02 \\
\hline\hline
\end{tabular}
\end{center}
\end{table}

\begin{figure*}
    \includegraphics[width=2.1\columnwidth]{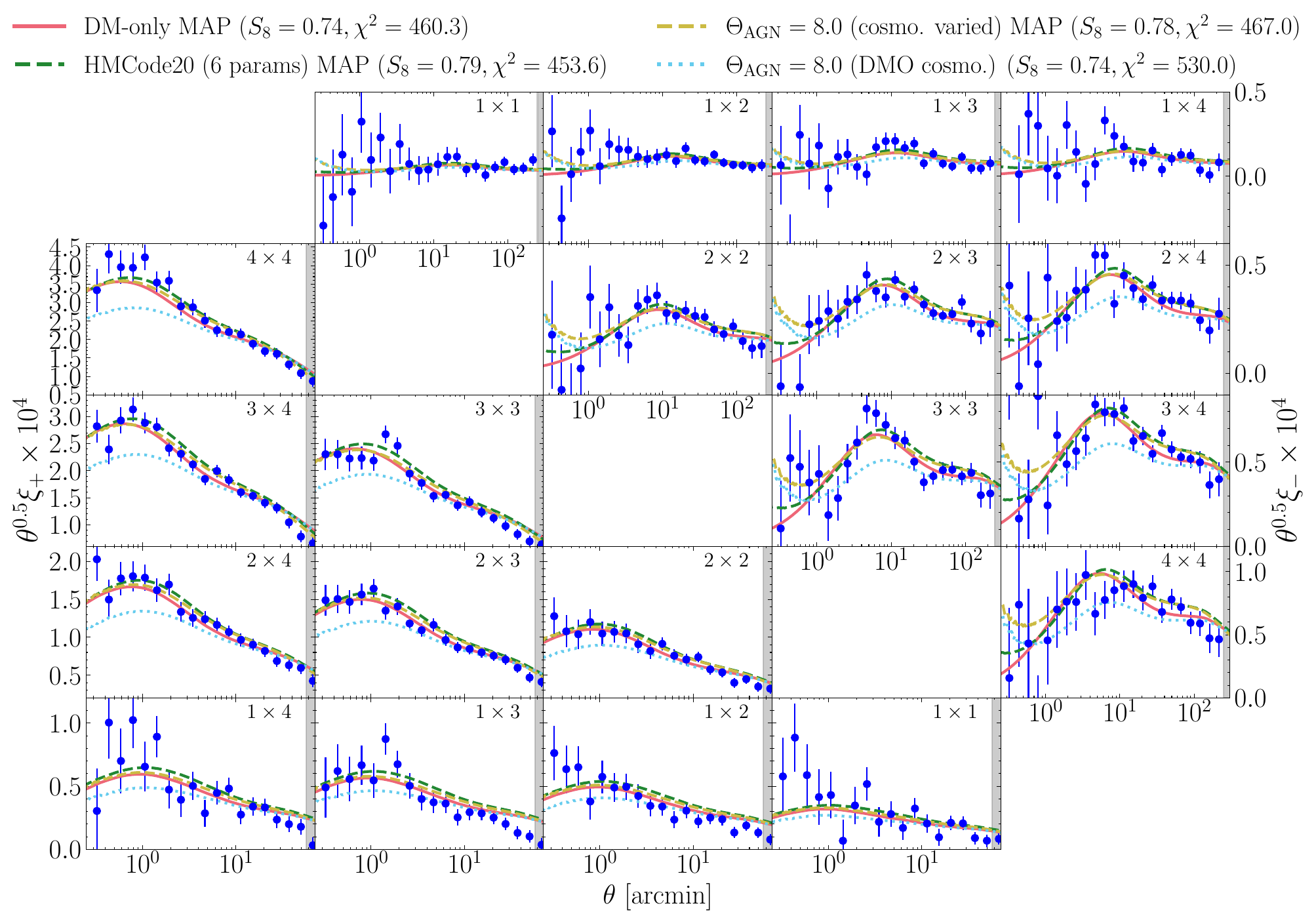}
    \caption{The ten 2PCFs including four auto-correlations and six cross-correlations between 
    the four tomographic bins as in Fig.~\ref{fig:xipm_k-modes}.
    The data and error bars are the measurements from the HSC-Y3 data. 
    The magenta-solid line in each panel is the MAP model in the chain that is obtained from our fiducial analysis using the smallest scale
    cut, ``$\theta_{+{\rm min}}:0'\!.28$'', as shown in Fig.~\ref{fig:2D}. For comparison, we show the model by the light-blue dotted line using the 
    \textsc{HMCode20}, where 
    we use the same model parameters as in the magenta line but include the baryonic effect with the fixed AGN feedback parameter of
    $\Theta_{\rm AGN}=8.0$. This model shows large deviations from the data at small angular scales. 
    The yellow-dashed line shows the results obtained 
    when fixing the AGN feedback parameter $\Theta_{\rm AGN}=8.0$ in the single-parameter 
    \textsc{HMCode20} model, but allowing other parameters to vary in the inference. Note that the \textsc{HMCode20} model does not include the DM-only model, even for very small $\Theta_{\rm AGN}$ values.
    The green dashed line shows the results when using the 6-parameter \textsc{HMCode20} model, which includes the DM-only model and can lead to both enhancement and suppression in the matter power spectrum
    by different combinations of the 6 parameters (see Section~\ref{sec:discussion} for details). 
    The legend also gives the central value of $S_8$
    and the $\chi^2$ values at the MAP model for each result. 
    }
    \label{fig:xipm_DM_mead20}
\end{figure*}

In this section, we evaluate the goodness-of-fit of the {\em maximum a posteriori} (MAP) model obtained from the fiducial chain, compared to the HSC-Y3 data. 
If the baryonic effect is significant 
in the small-scale cosmic shear signal, 
our method might not be able to fit the HSC-Y3 2PCFs at such small scales because the baryonic effect generally leads to different scale- and redshift-dependent changes in the 2PCFs from those due to changes in the cosmological parameters 
as illustrated in Fig.~\ref{fig:xipm}.

First, the magenta-solid lines in Fig.~\ref{fig:xipm_DM_mead20}
show the 2PCFs predictions of the MAP model obtained using the fiducial analysis method with the smallest scale cut (``$\theta_{+{\rm min}}:0'\!.28$''). The MAP model fairly well reproduces the measurements over the range of angular scales and in the different tomographic bins. For comparison, the light-blue, dotted lines show the model predictions computed using the \textsc{HMCode20}, where we used the same cosmological 
and nuisance parameters as those at the MAP model of the magenta line, but included the baryonic effect using the fixed AGN feedback parameter of 
$\Theta_{\rm AGN}=8.0$. 
This comparison model gives almost identical results to our model on large angular scales, but very different results on small scales where the baryonic effects are significant. It is interesting to note that the baryonic effect model predicts an upturn shape of $\xi_-$ at $\theta\lesssim 1$~arcmin, corresponding to the increase in the $P(k)$ amplitude at $k\gtrsim 20~h{\rm Mpc}^{-1}$ in Fig.~\ref{fig:Pk_ratio}.
To be quantitative, 
the $\chi^2$ value of this comparison model is worse than that of the MAP model by 
$\Delta \chi^2\simeq 70$ (more than 8$\sigma$), meaning that naively adding the baryonic effect to the model significantly degrades the goodness-of-fit. 
Of course, the changes in the 2PCFs due to the baryonic effect can be restored by changing other model parameters. Hence, the large $\chi^2$ value 
does not mean that the baryonic effect is strongly disfavored by the HSC-Y3 data, and we will study this in the next section. 
Note that the best-fit model slightly overestimates the 2PCFs on large scales. If we use the fiducial 
scale cut ``$\theta_{+{\rm min}}:7'\!.1$'' for our fiducial analysis setup using the DM-only model, 
the discrepancy disappears as can be seen in Fig.~3 of Li+HSCY3. The best-fit model of Fig.~\ref{fig:xipm_DM_mead20}
is mainly determined by the small-scale 2PCFs which have a higher signal-to-noise ratio.
Therefore, the subtle discrepancy on large scales could be due to the scale-dependent effect, which is not described by the DM-only model. However, this is beyond the scope of this paper, and will be studied
 in a separate paper.

Now we evaluate the goodness-of-fit with the $\chi^2$ value at MAP.
Since many of the parameters are prior dominated (also see Section~VA and Appendix~A of Li+HSCY3 for the similar discussion), the calculation of the number of degrees of freedom is not straightforward. Hence, we use noisy mocks of 2PCFs simulated according to the covariance matrix for the goodness-of-fit estimation. 
Noises with different realizations are added to the simulated data vector according to the covariance matrix. 
Note that, for the simulated data vector,  
we use the cosmological parameters and the nuisance parameters at the MAP model for the ``$\theta_{+{\rm min}}:0'\!.28$'' scale cut. 
We analyze these 50 mocks using our fiducial analysis method for each scale cut case
and, to save computational time, we sample them with \textsc{MultiNest}. We obtain the reference $\chi^2$ distribution from the histogram of the MAP values estimated from the 50 \textsc{MultiNest} chains.

Table~\ref{tab:goodness-of-fit} gives the $\chi^2$ value at MAP as well as the $p$-value. 
Here we estimate the $p$-value by computing the probability in the reference $\chi^2$ distribution for having higher $\chi^2$ values than the observed value ($\chi^2_{\rm MAP}$) of the data. 
The $p$-values exceed $0.1$ for all the different scale cuts except the smaller two scale cuts (``$\theta_{+{\rm min}}:1'\!.2$'' and ``$\theta_{+{\rm min}}:0'\!.28$''). The $p\simeq 0.02~ (0.03)$
for the ``$\theta_{+{\rm min}}:0'\!.28$'' (``$\theta_{+{\rm min}}:1'\!.2$'') scale cut is acceptable, but small. The small value might be due to an unknown signature at small scales such as the baryonic effect that is not well described by the DM-only model. However, this is not yet conclusive, and we instead conclude that the DM-only model can give an acceptable fit to the HSC-Y3 data over the range of scales down to $0.28$~arcmin we considered. We will below discuss the results for the different analysis setups; 
one includes IA (denoted ``+IA'' in Table~\ref{tab:goodness-of-fit}) and the other applies uninformative prior on $\Delta z_3$ and $\Delta z_4$ (denoted ``$\mathcal{U}(\Delta z_{3,4})$'' in Table~\ref{tab:goodness-of-fit}).

\subsection{A significance estimate of almost no shift in $S_8$ for the different scale cuts}
\label{subsec:S8shift}

\begin{figure*}
    \includegraphics[width=1.8\columnwidth]{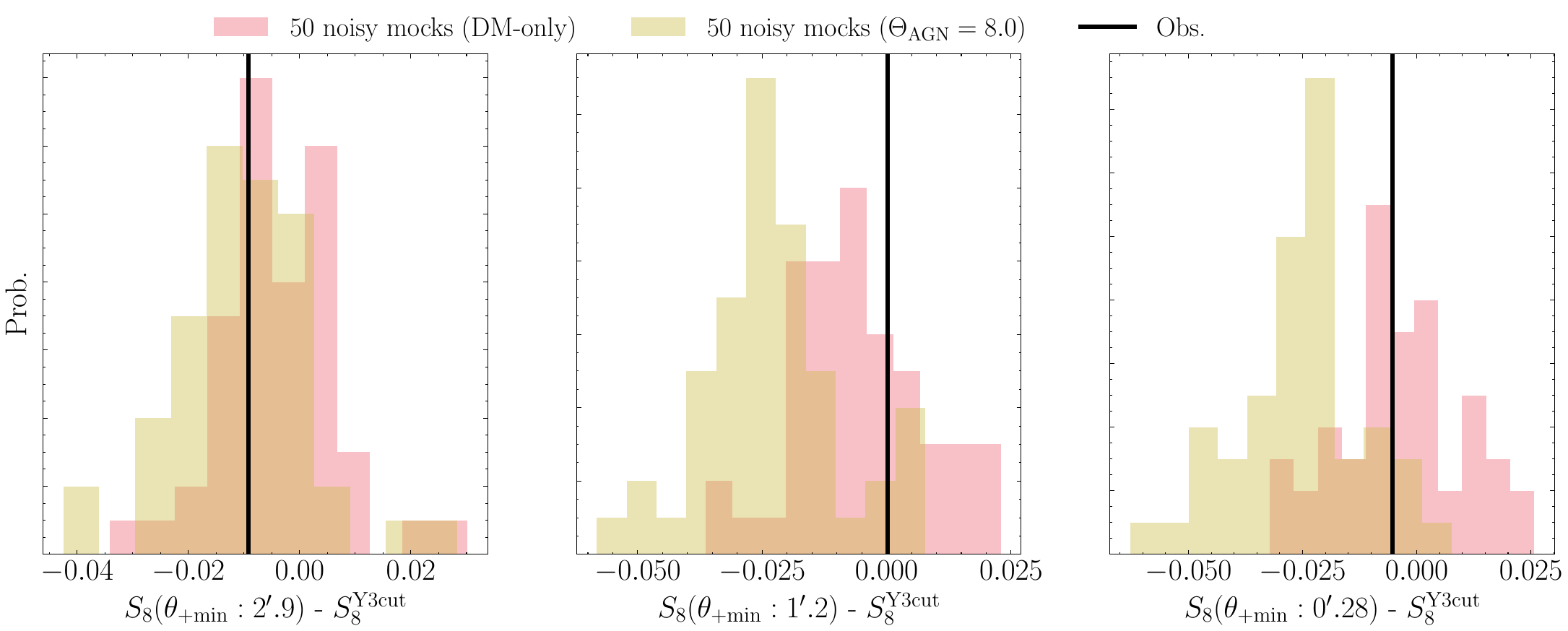}
    \caption{The magenta-shaded histogram shows the expected distribution of the differences between the 
    best-fit $S_8$ values obtained when using the different scale cut, compared to the Li+HSCY3 cut 
    (``$\theta_{+{\rm min}}:7'\!.1$''), named as $S_8^{\rm Y3cut}$. We generate the 50 noisy mock data around the simulated signal
    using the MAP model for our fiducial analysis method using the smallest scale cut (``$\theta_{+{\rm min}}:0'\!.28$''), and then analyze 
    each of the 50 noisy mock data with our fiducial analysis method. The histogram is the distribution of the 
    $S_8$ differences obtained from the hypothetical 50 analyses. 
    The yellow-shaded histogram shows the similar results, but using the 50 noisy mock data around the simulated signal at the MAP model, obtained using the \textsc{HMCode20} model with 
    the fixed $\Theta_{\rm AGN}=8.0$ (yellow-color dashed line in Fig.~\ref{fig:xipm_DM_mead20}). 
    Note that we use the same seed for each of the 50 noisy mock realizations. The vertical bold solid lines are the observed value of the $S_8$ difference from the HSC-Y3 data 
    (i.e. the values computed from the ``$\theta_{+{\rm min}}$'' row of
    Fig.~\ref{fig:1D_IA}).}
    \label{fig:DeltaS8}
\end{figure*}

As shown in Figs.~\ref{fig:2D} and \ref{fig:1D_IA}, we did not find any significant shift in $S_8$ even when using the smallest scale cut (``$\theta_{+{\rm min}}:0'\!.28$''), compared to the fiducial cut (``$\theta_{+{\rm min}}:7'\!.1$'')  used in Li+HSCY3. This implies that the HSC-Y3 cosmic shear data are unlikely to be affected by the AGN feedback at the level of 
$\Theta_{\rm AGN}=8.0$, for which we expect the shift in $S_8$ as in Fig.~\ref{fig:2D_baryon}.

To give a more quantitative answer, we construct two types of 50 noisy mock data. 
One is the 50 noisy mock realizations, generated in the same way described in the previous section,
around the simulated signal based on the MAP model 
for the cosmological parameters and the nuisance parameters for our fiducial analysis using 
 ``$\theta_{+{\rm min}}:0'\!.28$'' scale cut. Here we use the DM-only \textsc{DarkEmulator2} model to compute the 
 simulated signal. 
The other is the 50 noisy mock data around the simulated signal computed using the \textsc{HMCode20} with the fixed AGN feedback parameter of $\Theta_{\rm AGN}=8.0$, {which is the same noiseless mock data vector as
used in Fig.~\ref{fig:2D_TAGN8_noiselessmock}}. 
We use the same seed to generate each realization of the 50 noisy mocks for the DM-only case and the $\Theta_{\rm AGN}=8.0$ case. 
We then analyze each of the 50 noisy mocks with our fiducial analysis method using the DM-only \textsc{DarkEmulator2} model. 

Fig.~\ref{fig:DeltaS8} shows the shifts in $S_8$ compared to the fiducial scale cut, $S_8(\theta_{+{\rm min}}) - S_8^{\rm Y3cut}$. When using the smaller scale cut, the magenta-color histogram shifts towards a negative value of $S_8(\theta_{+{\rm min}}) - S_8^{\rm Y3cut}$. This means that, if the HSC data is affected by the baryonic effect of $\Theta_{\rm AGN}=8.0$, the analysis using 
the DM-only model tends to underestimate the $S_8$ value because of the stronger suppression due to the baryonic effect on smaller scales. 
However, the $S_8$ shift is not significant compared to the width of the histogram. 
We find the shifts in $S_8$ observed from the HSC data are consistent with both the mock results.  
Hence, for the current statistical errors of the HSC-Y3 data, we cannot distinguish the baryonic effect {from the shifts in $S_8$}
at a high significance {\em if} the effect is at the level of $\Theta_{\rm AGN}=8.0$.

\section{Discussion}
\label{sec:discussion}

For the parameter inference of the HSC-Y3 cosmic shear measurements, we have so far used 
the DM-only \textsc{DarkEmulator2} model to compute the model template. Although the DM-only model 
{is only an approximation in the sense that}
we have ignored the IA and baryonic effects, we have not found any clear failure point of the model in the parameter inference to within the statistical errors. 
In this section, we study the impact of the effects we ignored in our analyses so far. 

\subsection{A forward modeling of the baryonic effect}
\label{subsec:baryon_constraints}

\begin{table}
\caption{
Model parameters and priors of \textsc{HMCode20}, which models the matter power spectrum including the baryonic effect. We use two models of \textsc{HMCode20}: one is 
``6-parameter model'' that models the baryonic effect
by six parameters, and the other is the ``1-parameter model''
that uses only one model parameter, $\Theta_{\rm AGN}$, motivated by the fact that the AGN feedback is a dominant effect on scales relevant to the cosmic shear signal (see Section~\ref{subsec:baryon_constraints} for details).
}
\label{tab:feedback_parameters}
\setlength{\tabcolsep}{20pt}
\begin{center}
\begin{tabular}{ll}  \hline\hline
Parameter & Prior \\ \hline
\multicolumn{2}{c}{\bf 6-parameter model 
}
\\
$B_0$                               & ${\cal U}(3.13, 4.00)$ \\
$B_z$                            & ${\cal U}(-0.075, -0.039)$ \\
$f_{*,0}$                               & ${\cal U}(0, 0.0265)$ \\
$f_{*,z}$                            & ${\cal U}(0.39, 0.42)$ \\
$\log_{10}(M_{{\rm b}, 0} / h^{-1} M_{\odot})$                     & ${\cal U}(0, 14.83)$ \\
$M_{{\rm b}, z}$                     & ${\cal U}(-0.16, 0.57)$ \\
\hline 
\multicolumn{2}{c}{\bf 1-parameter ($\Theta_{\rm AGN}$) model}\\ 
$\Theta_{\rm AGN}$ & ${\cal U}(7.3, 8.3)$ 
\\ 
\hline
\hline
\end{tabular}
\end{center}
\end{table}

\begin{figure}
    \includegraphics[width=\columnwidth]{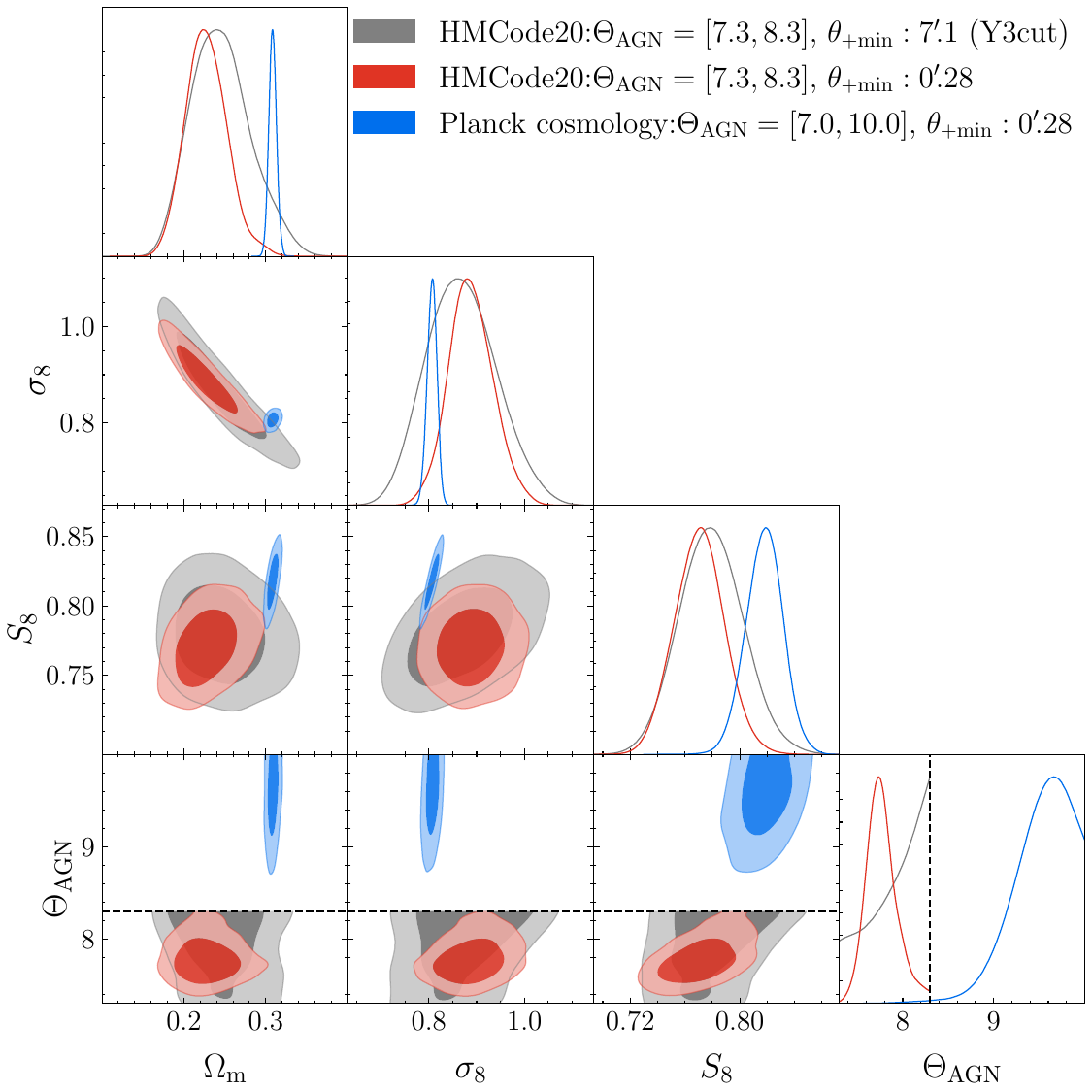}   
    \caption{The marginalized 1D and 2D posteriors of analysis using the single-parameter ($\Theta_{\rm AGN}$) \textsc{HMCode2020} model to compute the model template.
    The gray and red shaded regions show the results when using the prior range of $\Theta_{\rm AGN}=[7.3,8.3]$ as recommended 
    in \citet{halofit_mead21}, for the scale cuts of ``$\theta_{+{\rm min}}:7'\!.1$'' (Y3cut) and ``$\theta_{+{\rm min}}:0'\!.28$'', respectively.
    The blue regions show the results when we allow the prior range to outside the supporting range as given by $\Theta_{\rm AGN}=[7.0,10.0]$, and when we impose 
    the Gaussian prior on each of the cosmological parameters inferred from the {\it Planck} cosmology \citep{PlanckPR4} (see text for details).
    }
    \label{fig:2D_baryon}
\end{figure}

\begin{figure*}
    \includegraphics[width=1.8\columnwidth]{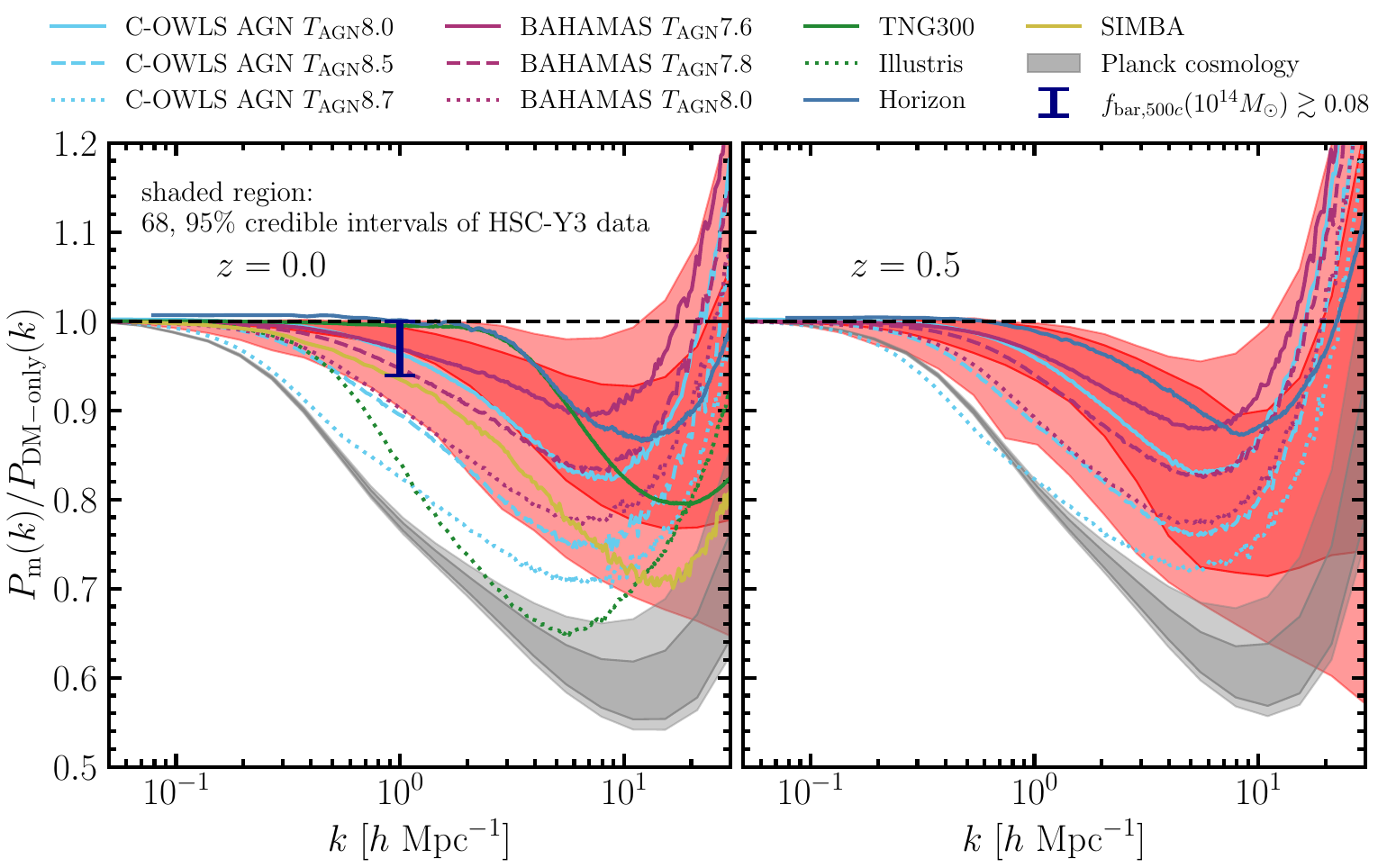}
    \caption{The dark and light red-color shaded regions are the 68\% and 95\% credible intervals of the model predictions in each $k$ bin, which are computed from the posterior distribution of the parameter inference using the 6-parameter \textsc{HMCode20} model for the ``$\theta_{+{\rm min}}:0'\!.28$'' scale cut (see text for details). The left and right panels show the results at $z=0$ and 0.5, respectively. For comparison, we show the results for the cosmological hydrodynamical simulations, \textsc{COSMO-OWLS} \citep{cowls_LeBrun2014}, 
    \textsc{BAHAMAS} \citep{2018MNRAS.476.2999M}, \textsc{Horizon} \citep{2014MNRAS.444.1453D}, \textsc{SIMBA} \citep{2019MNRAS.486.2827D},
    \textsc{Illustris} \citep{Illustris_Vogelsberger2014}, and \textsc{Illustris-TNG300} \citep{2018MNRAS.475..676S}, respectively. 
    {For $z=0.5$, we plot only those simulations that are available at the redshift. 
    The error bar in the plot of $z=0.0$ denotes a lower limit of the suppression indicated by the baryon mass fraction of cluster-scale halos,
    given as $f_{\rm bar,500c}(10^{14}M_\odot)\gtrsim 0.08$,  
    based on the observations such as the X-ray data \citep{2020MNRAS.491.2424V,FLAMINGO_project,HSC-XXL22} (see text for details).}
    The gray shaded regions are the similar results, which are computed from the posterior distribution in the chain for the result labeled ``+Planck 
    cosmology'' in Fig.~\ref{fig:2D_baryon}. 
    The $\chi^2$ value at the MAP model for the HSC-Y3 cosmic shear 2PCFs is degraded 
    by about $3\sigma$, compared to the value at the MAP model for our fiducial analysis result or the 6-parameter \textsc{HMCode20} result, but we here show the results for comparison.
    }
    \label{fig:Pk_ratio_constraints}
\end{figure*}

An alternative approach to our method is to use a forward modeling method of the baryonic effect
in the parameter inference. However, the drawback of such a method is that there are still large uncertainties in modeling the nonlinear matter power spectrum including the baryonic effects: different hydrodynamical simulations predict quite different large-scale structure formation on scales relevant to cosmic shear cosmology, 
depending on which recipes of the baryonic physics are used \citep[e.g.][]{2019OJAp....2E...4C}. 
Nevertheless, it would be worth investigating whether the use of a reference model that includes the baryonic effect can fit the HSC-Y3 data. 

For the purpose mentioned above, we use \textsc{HMCode20} \citep{halofit_mead21}, which models the cosmological dependence of and the baryonic effect on the matter power spectrum as a function of $k$ and redshift.  
The model describes clustering of three components: dark matter, stellar component that approximates a central galaxy by a point mass in its host halo, and diffuse intra-halo gas. Then \textsc{HMCode20} models relative fractions of the central galaxy and the diffuse gas in the total budget of baryonic matter, and models variations in the radial profile of total matter (dark matter plus the diffuse gas) in halos by changes in the halo concentration, from the DM-only Navarro-Frenk-White profile~\citep{NFW}, as a function of halo mass. \textsc{HMCode20} uses 6 parameters as given in Table~\ref{tab:feedback_parameters}: the stellar mass fraction ($f_{\ast,0}$); the characteristic halo mass scale $M_{b,0}$ below which diffuse gas is blown away from the host halo by the baryonic feedback effect and above which intra-halo diffuse gas is confined in the gravitational potential well of the host halo; and the normalization parameter $B_0$ for the halo mass concentration. And it has the three additional parameters to describe the redshift dependence of the three parameters. Changing these parameters can reproduce the DM-only model and can
describe both enhancement and suppression in the matter power spectrum amplitudes, which physically correspond to the baryon contraction \citep[e.g.][]{2004ApJ...616...16G} and the baryonic feedback effects of supernovae and AGNs \citep[e.g.,][]{2003ApJ...599...38B,2015JCAP...12..049S,baronCorrect_Schneider2019}, respectively.
Among these parameters, the halo concentration and the diffuse gas contribution are most relevant for the cosmic shear signal on scales we are interested in. 
\citet{halofit_mead21}
calibrated the model parameters using different data sets of the fitting formulae and the 
cosmological $N$-body and hydrodynamical simulations.
In particular, for the parameters describing the baryonic effects, they used data from the power spectrum library\footnote{https://powerlib.strw.leidenuniv.nl} of \citet{2020MNRAS.491.2424V} which contains simulations from the 
\textsc{COSMO-OWLS} \cite{cowls_LeBrun2014}
and \textsc{BAHAMAS} \cite{2018MNRAS.476.2999M} suites. 
Furthermore, \citet{halofit_mead21} proposed an effective model of the baryonic effect given by a single parameter, $\Theta_{\rm AGN}=\log_{10}(T_{\rm AGN}/{\rm K})$, motivated by the fact that the AGN feedback is the most important on the scales relevant to the cosmic shear signal and $T_{\rm AGN}$ roughly corresponds to the heating temperature due to the AGN feedback, which is responsible for the expansion effect of gas from the host halos. \citet{halofit_mead21} showed that 
variations in the 6 parameters found from a calibration set of the hydrodynamical simulations are fairly well captured by changes in $T_{\rm AGN}$ \citep[see Table~5 of][]{halofit_mead21}.
However, note that the single-parameter ($\Theta_{\rm AGN}$) \textsc{HMCode20} model does not include the DM-only model; even for small $\Theta_{\rm AGN}$ values, the model still predicts that the AGN feedback is not efficient in suppressing star formation, and therefore
the baryonic concentration is more effective, leading to an increase in the amplitude of matter power spectrum
at small scales compared to the DM-only model.

Thus, \textsc{HMCode20} is an empirical model, but was developed on the basis of the physical picture of the baryonic effects in structure formation. We use \textsc{HMCode20}, instead of the DM-only \textsc{DarkEmulator2}, to model the matter power spectrum in the parameter inference of the HSC-Y3 cosmic shear 2PCFs. We use both the \textsc{HMCode20} models using the single parameter ($\Theta_{\rm AGN}$) and the 6 parameters in Table~\ref{tab:feedback_parameters}. The priors on these parameters are given in Table~\ref{tab:feedback_parameters}, and the range of the priors corresponds to the range of the baryonic effects that are covered by the hydrodynamical simulations used in \citet{halofit_mead21}. In addition to these baryon parameters, we use the same cosmological and nuisance parameters and the same priors in Table~\ref{tab:parameters} 
as used in our fiducial analysis. 
Again note that the 6-parameter \textsc{HMCode20} model includes the DM-only-like model of matter power spectrum: the model with a specific set of the 6 
parameters given in Table~6 of \citet{halofit_mead21} corresponds to the DM-only model.

The yellow-dashed and green-dashed lines in Fig.~\ref{fig:xipm_DM_mead20} show the predictions at MAP in the chains obtained using the 6-parameter and 1-parameter \textsc{HMCode20} models for the ``$\theta_{+{\rm min}}:0'\!.28$'' scale cut, respectively. Note that the result for the 1-parameter ($\Theta_{\rm AGN}$) \textsc{HMCode20} is obtained by fixing $\Theta_{\rm AGN}=8.0$, but allowing the other parameters to vary in the parameter inference. 
We also note that the best-fit model prediction 
obtained by also allowing 
$\Theta_{\rm AGN}$ to vary is almost indistinguishable from the best-fit prediction obtained by fixing $\Theta_{\rm AGN}=8.0$.
Both the best-fit \textsc{HMCode20} models give very similar predictions to that of the best-fit DM-only model (the magenta line). The 6-parameter \textsc{HMCode20} model has a smaller $\chi^2$ value than that of the 1-parameter \textsc{HMCode20} model, reflecting that the 6-parameter model is more flexible. It is interesting to note that the best-fit 6-parameter model does not have an upturn shape in $\xi_-$ on small scales of $\theta\lesssim $~a few~arcmin similarly to the DM-only model, which is the main reason for the smaller $\chi^2$ value than does the 1-parameter model.

The dark and light gray-color shaded regions in Fig.~\ref{fig:2D_baryon} show the posteriors for the parameters 
$(\Omega_{\rm m}$, $\sigma_8$, $S_8$, $\Theta_{\rm AGN}$) for the analysis using the 1-parameter 
($\Theta_{\rm AGN}$)
\textsc{HMCode20} model, where 
we also allow $\Theta_{\rm AGN}$ to vary in the parameter inference. To do this, we impose the prior range on $\Theta_{\rm AGN}$ as given 
by $\Theta_{\rm AGN}=[7.3,8.3]$ as recommended by \citet{halofit_mead21}.  
The figure shows that, if we use the HSC-Y3 cosmic shear 2PCFs down to the smallest scale
(``$\theta_{+{\rm min}}:0'\!.28$''), 
the AGN feedback parameter $\Theta_{\rm AGN}$ is constrained, pointing {$\Theta_{\rm AGN}\simeq 7.8$}.
The MAP model in the chain has {$\chi^2\simeq 465.6$ and $S_8\simeq 0.773$}. These numbers are not so different from the result in Fig.~\ref{fig:xipm_DM_mead20}, where we used the fixed $\Theta_{\rm AGN}=8.0$ in the inference.  

In Fig.~\ref{fig:Pk_ratio_constraints},
the red-color shaded regions show the 68\% and 95\% credible intervals of the suppression in the matter power spectrum in each $k$ bin at
 $z=0$ and $z=0.5$, which are computed using the posterior for the analysis using the 6-parameter 
\textsc{HMCode20} for the ``$\theta_{+{\rm min}}:0'\!.28$'' scale cut case. The posterior indicated by the HSC-Y3 cosmic shear data does not prefer too strong suppression in the power spectrum amplitude. To be more quantitative, the power spectrum having the suppression up to $\sim 5\%$ at $k=1~h{\rm Mpc}^{-1}$ is consistent with the HSC-Y3 cosmic shear data. Our results are consistent with some of the hydro-simulations such as 
\textsc{Cosmo-OWLS} with $T_{\rm AGN}=8.0$, \textsc{BAHAMAS} with $T_{\rm AGN}=7.6$, \textsc{Horizon}
and \textsc{Illustris-TNG300}. We should also note that the DM-only model, which corresponds to unity in the ratio ($y$ axis), is consistent with the HSC-Y3 data. The 6-parameter \textsc{HMCode20} model 
tends to predict suppression (below unity in the $y$-axis) for most of the parameters range, in particular 
at $k\lesssim $~a~few~$h{\rm Mpc}^{-1}$. Hence, the ratio greater than unity is not necessarily disfavored, and the shaded region is due to the use of the restrictive model of matter power spectrum. Nevertheless, it would be worth noting that the model having the power greater than unity in this plot would correspond to even smaller $S_8$ value when fitted to the HSC-Y3 data, 
which increases the $S_8$ tension.

Finally, we study how much suppression in the matter power spectrum is required to
make the HSC-Y3 cosmic shear data consistent with the {\it Planck} CMB-inferred cosmology. 
To do this, we first adopt the priors on the cosmological parameters that are consistent with the {\it Planck}
$\Lambda$CDM model from \citet{PlanckPR4}. To be precise, we simply adopt the Gaussian prior for each parameter, which is 
taken from Table~5 of \citet{PlanckPR4}: 
${\cal N}(0.3092,0.0070)$ for $\Omega_{\rm m}$, ${\cal N}(2.091,0.029)$ for $10^9\times A_{\rm s}$,
${\cal N}(0.9681,0.0039)$ for $n_{\rm s}$, ${\cal N}(0.6764,0.0052)$ for $h$, and ${\cal N}(0.02226,0.00013)$
for $\omega_{\rm b}$, respectively. These correspond to much tighter constraints on the cosmological parameters than those from the HSC-Y3 cosmic shear data. 
To model the {\em extreme} AGN feedback effect, 
we use an ``extended'' model of the 1-parameter \textsc{HMCode20} model in the sense that 
we allow the model parameter $\Theta_{\rm AGN}$ to have larger values outside the calibrated range of $\Theta_{\rm AGN}$ (up to $\Theta_{\rm AGN}=8.3$)
in \citet{halofit_mead21}~\citep[e.g.][for similar approach]{Amon&Efstathiou22}. 
Extrapolation is dangerous, but the 1-parameter \textsc{HMCode20} model with such a large $\Theta_{\rm AGN}$ effectively predicts an extremely large suppression in the matter power spectrum.
The blue shaded regions in Fig.~\ref{fig:2D_baryon} show the posteriors obtained from the parameter inference when using the ``$\theta_{+{\rm min}}=0'\!.28$'' scale cut. 
Since the cosmological constraints 
from the {\it Planck} CMB data are so tight, the extreme baryonic effect, with {$\Theta_{\rm AGN}\simeq 9.8$}, can fit the HSC-Y3 cosmic shear 2PCFs. To be more precise, we find {$\Omega_{\rm m} = 0.311$, $\sigma_8 = 0.811$, $S_8 = 0.825$ 
and $\Theta_{\rm AGN} = 9.83$}, respectively at MAP. The MAP model is acceptable, but has a larger 
$\chi^2$ value 
than does the MAP model of the DM-only model in Fig.~\ref{fig:xipm_DM_mead20} by {$\Delta \chi^2\simeq 9.7$}, therefore about $3\sigma$ degradation. 

The gray shaded regions in Fig.~\ref{fig:Pk_ratio_constraints} denotes the suppression in 
$P_{\rm m}(k)$ when the cosmology model is fixed to the {\it Planck} $\Lambda$CDM model {(using the Gaussian prior)} and when the extreme baryonic suppression is allowed. The suppression inferred from this analysis is larger than indicated by any of the existing hydrodynamical simulations; it requires about 25\% fractional suppression at $k\sim 
1~h{\rm Mpc}^{-1}$\footnote{Note that the suppression in the original \textsc{Illustris} simulation \citep{Illustris_Vogelsberger2014} is too large and was found to be due to the unphysical errors in the numerical implementation of AGN feedback.}. However, this level of suppression is disfavored
by other observations as described in the following.
{\citet{2020MNRAS.491.2424V} found that 
a scaling relation between the baryonic suppression amplitude 
in $P_{\rm m}(k)$ at $z = 0$ at $k \sim 1.0 ~h {\rm Mpc}^{-1}$ and the mean baryon mass fraction in cluster-scale ($\sim 10^{14} M_{\odot}$) halos (denoted as ${f}_{{\rm bar}, 500c}(10^{14} M_{\odot})$) normalized by the cosmic baryon fraction, $\tilde{f}_{{\rm bar}, 500c}(10^{14} M_{\odot}) \equiv {f}_{{\rm bar}, 500c}(10^{14} M_{\odot})/(\Omega_{\rm b}/\Omega_{\rm m})$, holds for most of the hydrodynamical simulations (also see Fig.~23 of \citep{FLAMINGO_project} that includes recent simulations). The 
vertical bar at $k=1~h{\rm Mpc}^{-1}$ in the $z=0$ panel of Fig.~\ref{fig:Pk_ratio_constraints} 
denotes up to $\sim 7\%$ suppression at $k = 1~h{\rm Mpc}^{-1}$ that is indicated by the lower bound of the baryon mass 
fraction based on the $X$-ray and weak lensing observations for a sample of clusters, given by $f_{\rm bar, 500c}(10^{14}M_\odot)\gtrsim 0.08$ corresponding to $\tilde{f}_{\rm bar,500c}(10^{14}M_\odot)\gtrsim 0.5$ since $\Omega_{\rm b}/\Omega_{\rm m}\simeq 0.157$.  
In particular, \citet{HSC-XXL22} combined the X-ray, optical and weak lensing measurements for a sample of clusters to 
estimate the masses of diffuse thermal gas, stellar mass (mainly BCG's stellar mass) and total matter (mainly dark matter), arriving at an estimate 
of $\tilde{f}_{\rm bar, 500c}\simeq 0.5$--$0.7$ (see Fig.~5 of \citep{HSC-XXL22}). 
Hence we conclude the 25\% fractional suppression at $k\simeq 1~h{\rm Mpc}^{-1}$ required by the 
{\it Planck} cosmology (the gray shaded region in Fig.~\ref{fig:Pk_ratio_constraints})
is strongly disfavored by these observational studies. 
}
{On the other hand, 
the red-color shaded region is consistent with the aforementioned observations, i.e. the lower bound denoted by the vertical bar.
This level of the suppression in the amplitude of $P(k)$ at $k=1~h{\rm Mpc}^{-1}$ is also consistent with the results using the DES Year~3 data 
\citep{baryon_chen22,baryon_arico23}.
The constraints from the KiDS-1000 cosmic shear data without combining with other probes \citep{KiDS1000_baryon_Schneider2021} are relatively weak and encompass our HSC-Y3 constraints.}

From these results, we conclude that it is unlikely the 
baryonic suppression  alone  resolves the $S_8$ tension in the HSC-Y3 data.

\subsection{IA effects}
\label{subsec:IA}
\begin{table}
\caption{
Model parameters and priors used in the TATT model of the IA effect. 
}
\label{tab:IA_parameters}
\setlength{\tabcolsep}{20pt}
\begin{center}
\begin{tabular}{ll}  \hline\hline
Parameter & Prior \\ \hline
\multicolumn{2}{l}{\hspace{-1em}\bf Intrinsic alignment parameters
(Section~\ref{subsec:IA})}\\
$A_1$                               & ${\cal U}(-6, 6)$ \\
$\eta_1$                            & ${\cal U}(-6, 6)$ \\
$A_2$                               & ${\cal U}(-6, 6)$ \\
$\eta_2$                            & ${\cal U}(-6, 6)$ \\
$b_\mathrm{ta}$                     & ${\cal U}(0, 2)$ \\
\hline
\hline
\end{tabular}
\end{center}
\end{table}
\begin{figure}
    \includegraphics[width=\columnwidth]{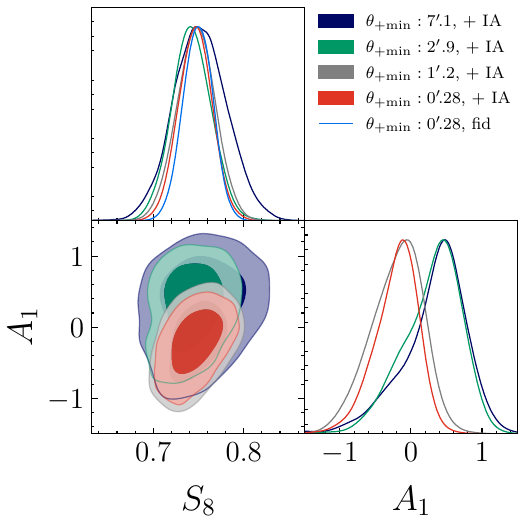}
    \caption{The marginalized 1D and 2D posteriors of the parameters $(S_8, A_1)$ when we include the IA effect 
    in the model template, for the different scale cuts. We use the TATT model of the IA effect, given by the 5 parameters and 
    the priors in Table~\ref{tab:IA_parameters}. The parameter $A_1$ is the linear IA coefficient.
    For comparison, we also show the 1D posterior of $S_8$ for our fiducial analysis with the scale cut ``$\theta_{+{\rm min}}:0'\!.28$'', which 
    is the same as in Fig.~\ref{fig:2D}. Note that our fiducial DM-only method does not include the IA modeling. 
    }
    \label{fig:2D_IA}
\end{figure}

So far, we have not included the IA effects in the theoretical template. This can be considered a conservative approach, because the leading-order contribution of 
IA effects, arising from the IA and cosmic shear cross-correlation \citep{ia_hirata_seljak04}, 
generally predicts a negative contribution to the cosmic shear 2PCFs \citep[e.g., see][and references therein]{2021MNRAS.501..833K,2021JCAP...03..030S}. This is in the same direction as the suppression of the baryonic effect, so ignoring the IA effect in the theoretical template would tend to 
give a worse fit if the HSC-Y3 cosmic shear data are affected by the IA and/or baryonic effects. 
Nevertheless, it is worth examining how the results change when we include the IA effect in the model template,
which is what this section aims to address. 

We use the tidal alignment and tidal torque model \citep[TATT;][]{tatt_blazek17}, which is the same model as used in Li+HSCY3. We use the same five parameters and the prior ranges as in Li+HSCY3 {(Table~\ref{tab:IA_parameters}).}
Then we use {Eq.~(14) in Li+HSCY3} to include the IA contribution to the total power of galaxy shape 2PCFs, and perform the Bayesian parameter inference. 
{Note that TATT model includes NLA~\citep{nla_bridle07, nla_hirata07} when $A_2 = 0$ and $b_{\rm ta} = 0$.}

As shown in the ``+IA'' row of Fig.~\ref{fig:1D_IA},
even if we include the IA effect in the model template, our results remain almost unchanged. 
Fig.~\ref{fig:2D_IA} shows the 1D and 2D posterior distributions when including the 
IA effect, for the different scale cuts, compared to the results for our fiducial analysis (i.e. the use of 
DM-only model). We didn't detect significant $|A_1| > 0$ for all 4 scale cuts. This suggests the data do not have a clear signature of the IA effects.

We also estimate the goodness of fit using the noisy mocks as in Section~\ref{subsec:goodness of fit}, but with IA effects. The results are listed in Table~\ref{tab:goodness-of-fit}. We found the DM-only model with the IA effects still can fit the data. 
Hence, our conclusion on the baryonic effects in the previous sections is not affected by the IA effects.

We note that the conclusion on the IA detection especially on small scales can be different for other IA models such as EFT model~\citep{ia_eft_bakx} and halo model~\citep{iaSDSS_Singh2015}. We will investigate the detectability of IA on small scales in future studies.

\subsection{The impact of the uninformative priors on $\Delta z_{3,4}$}
\label{sec:dz34}
\begin{figure}
    \includegraphics[width=\columnwidth]{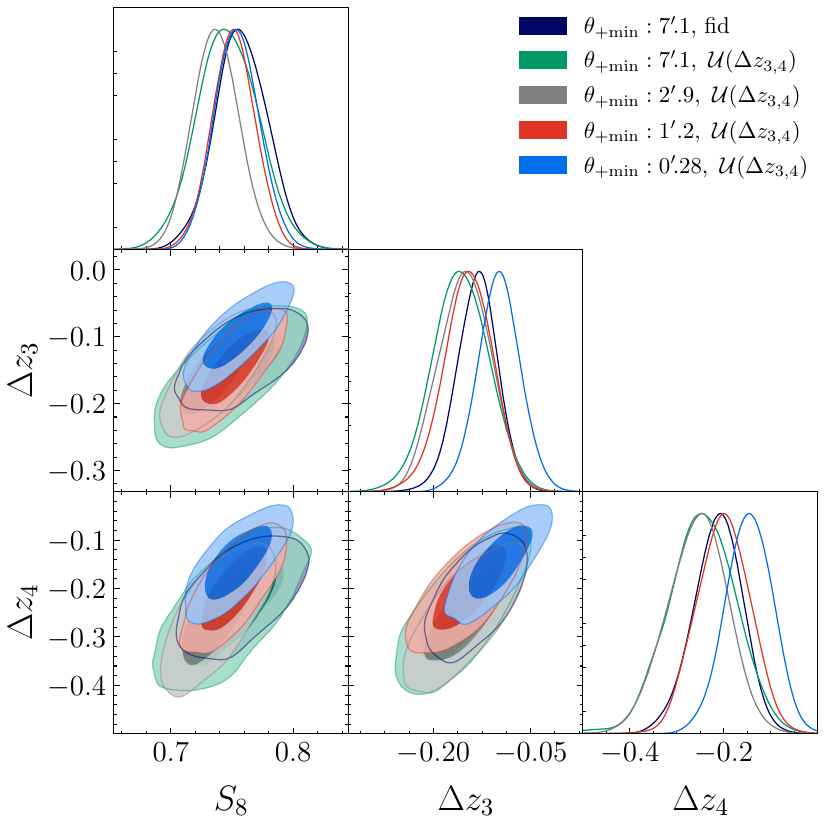}
    \caption{Similarly to Fig.~\ref{fig:2D_IA}, but this figure shows the posteriors when we use the uninformative flat priors on 
    $\Delta z_3$ and $\Delta z_4$, instead of the Gaussian prior in our fiducial analysis (Table~\ref{tab:parameters}).
    }
    \label{fig:2D_photoz}
\end{figure}

For our fiducial analysis, we adopt Gaussian informative priors on $\Delta z_3$ and $\Delta z_4$ in order for us
to focus on a possible signature of the baryonic effects in the parameter inference. 
In this section, we discuss how the use of uninformative priors 
on $\Delta z_3$ and $\Delta z_4$ change our conclusion.

The ``${\cal U}(\Delta z_{3,4})$'' row of Fig.~\ref{fig:1D_IA} shows that the results remain almost unchanged, 
compared to the results of our fiducial analysis with DM-only model. Note that, for these results, we did not 
include the IA effect in the model template. 
Fig.~\ref{fig:2D_photoz}  shows the 1D and 2D posterior distributions.

\subsection{The impact of reduced shear}
\label{sec:high-order_lensing}

\begin{figure*}
    \includegraphics[width=2.1\columnwidth]{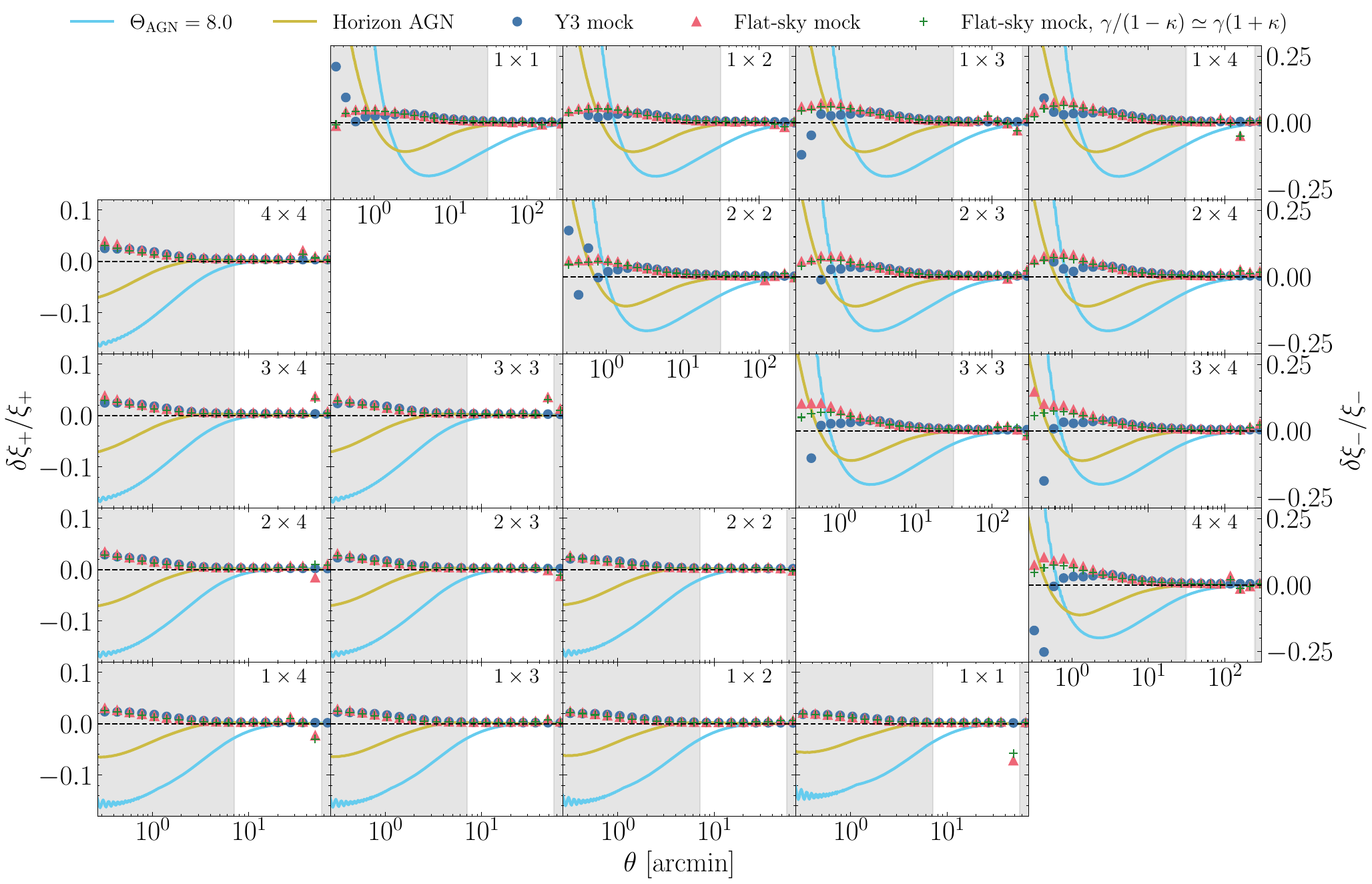}
    \caption{Each panel shows a fractional change in each of the tomographic cosmic shear 2PCFs
    due to the use of the reduced shear, 
    $g_\alpha=\gamma_\alpha/(1-\kappa)$, 
    compared to those computed using the weak lensing approximation 
    $g_\alpha\simeq \gamma_\alpha$. We use the mock HSC galaxy catalogs, where the weak lensing distortion effect 
    is simulated on each galaxy using the ray-tracing simulations (see text for details). 
    The unshaded region shows the scales used in the Li+HSCY3 analysis. 
    Red-triangle symbols in each panel show the result using the simulated reduced shear, while
    blue-circle symbols show the result 
        using the weak lensing approximation of $g_\alpha \simeq \gamma_\alpha$.
    Green-cross symbols show the result obtained when using the perturbative expansion as $g_\alpha\simeq 
    \gamma_\alpha(1+\kappa)$ in the mock catalog. For comparison, we also show the fractional change computed using the \textsc{HMCode20} with $\Theta_{\rm AGN}=8.0$ and the \textsc{Horizon} simulation result, compared 
    to the DM-only prediction for the fixed $\Lambda$CDM cosmology. 
        }
    \label{fig:thetaxi_RS}
\end{figure*}

\begin{figure*}
    \includegraphics[width=2.1\columnwidth]{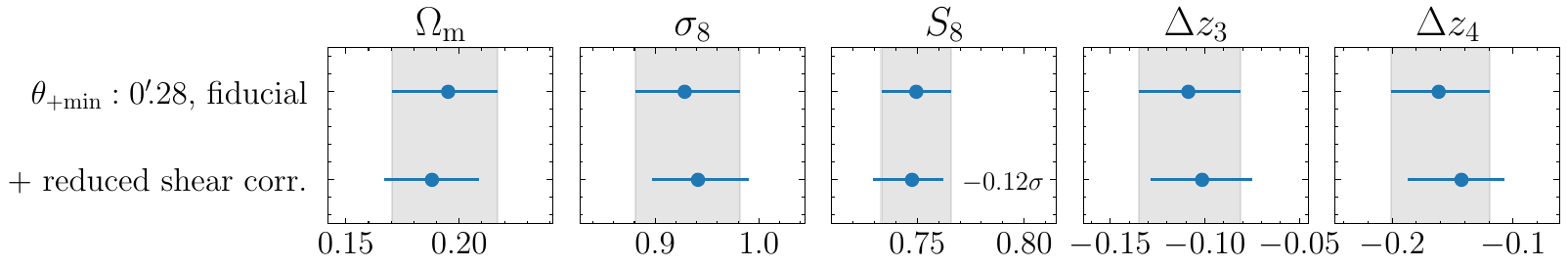}
    \caption{The impact of the reduced shear correction on the parameter inference, similarly to Fig.~\ref{fig:1D_IA}.
    We assume the {\em fixed} scale-dependent, fractional change (triangle symbols) in Fig.~\ref{fig:thetaxi_RS}
    and multiply the fractional change to each of the 2PCFs in the model template when varying cosmological models in the parameter inference. We apply this method to the HSC-Y3 cosmic shear 2PCFs data
    for the ``$\theta_{+{\rm min}}:0'\!.28$'' scale cut, 
    and the figure shows the central value and the 68\% credible interval for each parameter, compared to the results of our fiducial analysis method (the same as in the ``fiducial'' row of Fig.~\ref{fig:1D_IA}).
    The number above next to $S_8$ denotes the shift in the central value of $S_8$ with respect to the central value and the $1\sigma$ width of the fiducial analysis with the ``$\theta_{+{\rm min}}:0'\!.28$'' scale cut.
    }
    \label{fig:1D_reducedshear}
\end{figure*}

So far we have included the leading-order weak lensing effect 
for the theoretical template of cosmic shear 2PCFs. 
There are higher-order effects in the observed 2PCFs that might not be negligible on small scales:
the higher-order contributions include the post Born approximation, 
the lens-lens coupling 
\citep{2002ApJ...574...19C,2003ApJ...592..699V}, and the reduced shear \citep{2006PhRvD..73b3009D,2009ApJ...696..775S,2003MNRAS.346..949T}. 
The leading order ones are the post Born approximation and the reduced shear, 
while the lens-lens coupling and other effects are smaller. 
Here we consider the correction due to the reduced shear as a working example \citep{2006PhRvD..73b3009D,2009ApJ...696..775S,2003MNRAS.346..949T}. 
The reduced shear arises from the fact that the observed weak lensing distortion is due to the reduced shear, 
$g=\gamma/(1-\kappa)$, rather than the shear $\gamma$, where 
$\kappa$ is the convergence field, which is a projected surface mass density weighted by the lensing critical surface density for a given source redshift. Only in the weak lensing regime satisfying 
$|\gamma|, |\kappa|\ll 1$, we have $g\approx \gamma$, which we have so far assumed.

{Fig.~\ref{fig:thetaxi_RS} shows the magnitude of the reduced shear correction in each of the tomographic 
cosmic shear 2PCFs.
From the nature of the random variable ($\kappa$) in the denominator of $g$, it is difficult to analytically estimate the impact of the reduced shear. To estimate the amount of the correction, we use 
the ray-tracing simulations in \citet{raytracingTakahashi2017} (denoted as ``Y3 mock") and \citet{2021MNRAS.504.1825S} (denoted as ``Flat-sky mock")
for the $\Lambda$CDM model.}
First of all, the reduced shear correction is negligible on scales greater than 
the fiducial scale cut (``$\theta_{+{\rm min}}:7'\!.1$'') of Li+HSCY3. 
On the other hand, the correction appears on scales smaller than a few arcminutes, 
and the correction is larger for higher source redshifts. The correction amplitude is up to about {10\%} on the smallest scales. The magnitude of the correction is roughly comparable to the magnitude of the baryonic effect seen in 
\textsc{Horizon}, which predicts moderate suppression in the matter power spectrum. 
For $\xi_+$, the reduced shear correction is in the opposite direction to the baryonic suppression. For $\xi_-$, the correction could be in the same direction as the baryonic enhancement 
on the smallest scales. Finally, the perturbative approximation using the expansion 
$g\simeq \gamma(1+\kappa)$ is fairly good at these scales, and can be calculated analytically by using the matter bispectrum for a given cosmological model \citep{2006PhRvD..73b3009D,2009ApJ...696..775S,2003MNRAS.346..949T}.

Since the reduced shear correction should be present, we evaluate the impact on the parameter inference. 
Fig.~\ref{fig:1D_reducedshear} shows the result. Here we assume the ``fixed'' scale-dependent, fractional change in each of the cosmic shear 2PCFs and multiply the fractional change to each of the 2PCFs in the model template
when varying cosmologies. Fig.~\ref{fig:1D_reducedshear} shows the result when applying this method to the 
HSC-Y3 2PCFs data. The reduced correction causes a small but insignificant shift in $S_8$, in the direction of increasing the $S_8$ tension. Thus we can include the effect of the reduced shear in an iterative way 
assuming cosmologies around the best-fit model, if we want to include the small-scale information of 2PCFs  in the cosmology analysis. We conclude that the effect of the reduced shear does not change our main conclusion of this paper.

\subsection{Future prospects: the use of small-scale $\xi_-$ as a diagnostic of 
the baryonic effect}
\label{sec:prospects}
\begin{figure}
    \includegraphics[width=\columnwidth]{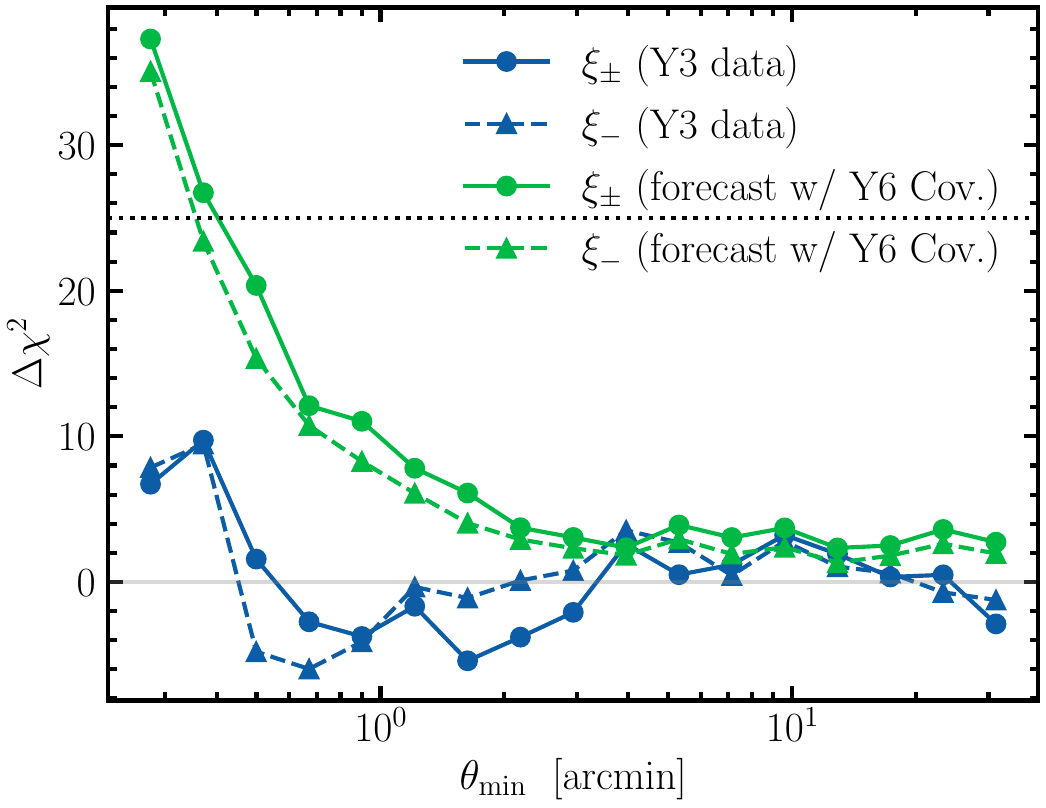}
    \caption{The blue data points show 
    the $\Delta \chi^2$ value 
    between the $\chi^2$
    values at the MAP models obtained when performing the parameter inference of the HSC-Y3 data with 
     either the DM-only model or the 1-parameter ($\Theta_{\rm AGN}$) \textsc{HMCode20} model with the fixed 
    $\Theta_{\rm AGN}=8.0$, $\Delta \chi^2\equiv \chi_{\rm MAP}^2(\mbox{\textsc{HMCode20}})-\chi^2_{\rm MAP}(\mbox{DM-only})$
    summed  over $\theta_{\rm min}\le\theta\le \theta_{\pm {\rm max}}$, {as a function of the chosen 
    small-scale cuts ($\theta_{{\rm min}}$) in the $x$-axis. 
    {Here 
    $\theta_{\pm{\rm max}}$ are fixed to the same as the maximum separations defined in Section~\ref{subsec:scale-cuts}.}
    The blue circle symbols show the result using both the $\xi_{\pm}$ data for the $\Delta\chi^2$ evaluation, 
    while the blue triangle symbols are calculated using the same MAP model as the circle symbols, 
    but using only the $\xi_{-}$ data. 
    The green circle and triangle symbols show the forecast for the HSC Year~6 data (i.e. the full data set covering about 1,100~sq.~degrees).
    To do this, we used the noiseless mock data computed using the MAP model of the 6-parameter \textsc{HMCode20} in Fig.~\ref{fig:xipm_DM_mead20}, then performed hypothetical analyses of the data using either the 6-parameter \textsc{HMCode20}
    model or the 1-parameter \textsc{HMCode20} model with $\Theta_{\rm AGN}=8.0$, for the different small-scale cuts.}
    In this case we used 
    $\Delta\chi^2\equiv \chi^2_{\rm MAP}(\mbox{1para-\textsc{HMCode20}})-\chi^2_{\rm MAP}(\mbox{6para-\textsc{HMCode20}})$.
     The dotted line denotes $\Delta \chi^2=25$ (about 
    $5\sigma$ threshold for the Gaussian distribution case) 
    for reference. 
    }
    \label{fig:Deltachi2}
\end{figure}

The $S_8$ constraint is mainly obtained by the $\xi_+$ information, as it carries a higher signal-to-noise ratio than $\xi_-$ does. On the other hand, $\xi_-$ at a given separation 
is more sensitive to the power spectrum at smaller scales,
including the baryonic effect, than is $\xi_+$. 
Hence, combining $\xi_+ $ and the small-scale $\xi_-$ can provide a useful way to constrain both the cosmological parameters
and the baryonic effect simultaneously.

In Fig.~\ref{fig:Deltachi2}, the blue solid and dashed lines show the $\Delta \chi^2$ calculated from the $\chi^2$ values at the MAP models obtained 
using the DM-only model or the 1-parameter \textsc{HMCode20} model with fixed $\Theta_{\rm AGN}=8.0$, {as a function of the small-scale cuts. The large-scale cuts ($\theta_{\pm{\rm max}}$)
are fixed to the same as in our fiducial analysis, and 
we impose the same small-scale cut for both $\xi_+$ and $\xi_-$; $\theta_{+{\rm min}} = \theta_{-{\rm min}} = \theta_{{\rm min}}$
 as given in the $x$-axis.}
Note that these two models have the same number of model parameters. 
This plot provides a metric to distinguish between the 
DM-only model and the \textsc{HMCode20} with $\Theta_{\rm AGN}=8.0$, given the measured $\xi_{\pm}$.
The figure shows that, for the HSC-Y3 cosmic shear 2PCFs, $\xi_-$ at $\theta\lesssim 1$~arcmin gives most of $\Delta \chi^2$, suggesting 
that $\xi_-$ can be a useful diagnostic of the baryonic effect on small scales. 

The green solid and dashed lines show the forecast of the $\Delta \chi^2$ diagnostic 
for the synthetic HSC Year~6 data assuming an area coverage of about 1,100~sq. degrees and 
the same depth and image quality as those in the HSC-Y3 data. To do this, we generate the noiseless mock data using the MAP model of the 6-parameter \textsc{HMCode20} in 
Fig.~\ref{fig:xipm_DM_mead20}, assuming that the MAP model reflects a realistic baryonic effect in the universe. We also properly scaled the covariance matrix to that for 1,100~sq.~degrees. 
{We plot the $\Delta\chi^2$ value between the MAP models for the \textsc{HMCode20} with $\Theta_{\rm AGN}=8.0$ and the 6-parameter \textsc{HMCode20}, where $\chi^2_{\rm MAP}$ for the latter model has a value of zero by construction. 
In this case, $\xi_-$ gives most of the contribution to $\Delta \chi^2$, especially on scales smaller than a few arcmin, and the baryonic effect represented by the \textsc{HMCode20} with $\Theta_{\rm AGN}=8.0$ is disfavored against the 6-parameter \textsc{HMCode20} baryon model, when including the small-scale information. }
A Stage-IV survey such as the Rubin Observatory LSST covering about 18,000~sq. degrees
should be able to more clearly distinguish the baryonic feedback models using the method proposed in this paper.

\section{Conclusion}
\label{sec:conclusion}

In this paper, we have used the cosmic shear 2PCFs measured from the HSC-Y3 data in four tomographic redshift bins, down to small angular separations $\theta=0.28$~arcmin, taking full advantage of the high number density of galaxies thanks to the deep HSC data. While the published HSC-Y3 cosmology analysis \citep[Li+HSCY3;][]{HSC3_cosmicShearReal} used the 2PCFs at scales, which are sensitive to the matter power spectrum at $k\lesssim 1~h{\rm Mpc}^{-1}$, the small-scale 2PCFs used in this paper allow us to explore a possible signature of the baryonic feedback effect in the nonlinear matter power spectrum 
down to $k\simeq 20~h{\rm Mpc}^{-1}$. 

For our purpose, we used an accurate model of the DM-only matter power spectrum, \textsc{DarkEmulator2}, calibrated 
from the high-resolution $N$-body simulation, to compute the theoretical templates of the cosmic shear 2PCFs used in the cosmology inference. The DM-only power spectrum model was found to be accurate to within one percent in the amplitude up to 
$k\sim 100~h{\rm Mpc}^{-1}$ for models around the HSC-Y3 inferred $\Lambda$CDM model. Our aim was therefore to minimize the modeling uncertainty. Our method differs from the conventionally used method, i.e. a forward modeling method in which the baryonic effects in the matter power spectrum are modeled by introducing a set of parameters to model the baryonic effects in an empirical way. The latter method suffers from uncertainties in the modeling of the baryonic effects as different hydrodynamical simulations give different predictions depending on which subgrid physics recipes are used. 
Our strategy was to test the working hypothesis that, if the HSC-Y3 cosmic shear data were affected by a significant baryonic effect, then the DM-only model would not be able to give a good fit of the data. 

Our findings are summarized as follows: 
\begin{enumerate}
    \item[$\bullet$] The DM-only model can well reproduce the HSC-Y3 cosmic shear 2PCFs down to 
    the smallest scale ($0.28$~arcmin). We evaluated the goodness-of-fit using the covariance matrix, and found 
    $p\simeq 0.02$ for the $p$-value when using the smallest scale cut. The small $p$-value could be an indication of unknown effects that cannot be accounted for by the DM-only model, but still represents an acceptable fit.  (Figs.~\ref{fig:2D}, \ref{fig:1D_IA} and \ref{fig:xipm_DM_mead20})
    \item[$\bullet$] The measured $S_8$ values show no significant shift when using the four different scale cuts
    from $7.1$ to $0.28$~arcmin, although a negative shift in the $S_8$ value would be expected if the significant suppression due to the baryonic effects were present in the data at the small scales. (Figs.~\ref{fig:2D} and \ref{fig:1D_IA})
    \item[$\bullet$] Using the noisy mock data, we showed that no apparent shift in $S_8$ is consistent with the statistical errors of the HSC-Y3 data, if the data follow the DM-only model predictions or 
    even if the data were affected by the AGN feedback model with 
    $\Theta_{\rm AGN}(\equiv \log(T_{\rm AGN}/{\rm K}))=8.0$, which corresponds to the largest feedback effect in the hydrodynamical simulations of \textsc{BAHAMAS}.
    (Fig.~\ref{fig:DeltaS8})
    \item[$\bullet$] Using the 6-parameter \textsc{HMCode20} model, which include the DM-only model and can lead to both enhancement and 
    suppression in the nonlinear matter power spectrum compared to the DM-only model by varying combinations of the 6 parameters, we showed that the HSC-Y3 2PCFs on the small scales are consistent with a fractional suppression of up to $5\%$ at $k\simeq 1~h{\rm Mpc}^{-1}$ or with a moderate magnitude of the baryonic feedback effect such as that seen in \textsc{Illustris-TNG} and \textsc{Horizon}. (Fig.~\ref{fig:Pk_ratio_constraints})
    \item[$\bullet$] To reconcile the HSC-Y3 2PCFs with the {\it Planck}-inferred $\Lambda$CDM model, it requires an extremely large suppression in the nonlinear matter power spectrum, such as a $25\%$ suppression 
    at $k\simeq 1~h{\rm Mpc}^{-1}$, corresponding to the AGN feedback with {$\Theta_{\rm AGN}\simeq 9.8$} which is much larger than the effect seen in any of the existing hydrodynamical simulations. 
    This amount of suppression is also strongly disfavored by the baryon mass fraction for cluster-scale halos
    indicated by the X-ray observations.
    (Fig.~\ref{fig:Pk_ratio_constraints})
    \item[$\bullet$] Our conclusion, no significant shift in $S_8$ or no failure point of the DM-only model, is unchanged even when we include the IA effect, employ the flat priors on the mean redshift errors of the third and fourth high-redshift bins, and consider the higher-order lensing effects such as the reduced shear. (Figs.~\ref{fig:1D_IA}, \ref{fig:2D_IA}, \ref{fig:2D_photoz}, and \ref{fig:1D_reducedshear})
\end{enumerate}

Thus we conclude that the $S_8$ tension, found in Li+HSCY3, is still present even when we perform the parameter inference of the HSC-Y3 cosmic shear 2PCFs down to the smallest scales. This result is consistent 
with the cosmology analysis using the HSC-Y3 cosmic shear power spectra \citep{HSC3_cosmicShearFourier} and with the $3\times$2pt cosmology analysis combining the HSC-Y3 weak lensing catalog and the SDSS DR11 spectroscopic catalog \citep{HSC3_3x2pt_ss}. We note that most of the cosmological information 
in the latter $3\times 2$pt analysis comes from the SDSS sample, which covers a much larger area of 8,000~sq. degrees, so the $S_8$ tension found there and in this paper can be considered almost independent.

Many other large-scale structure probes similarly indicate the $S_8$ tension: other weak lensing 
surveys \citep{DESY3_CS_Secco2022,DESY3_CS_Amon2021,KiDS1000_CS_Asgari2020},
the redshift-space galaxy clustering \citep{2020JCAP...05..042I,2022PhRvD.105h3517K,2022JCAP...02..008C},
and the tomographic cross-correlation of DESI luminous red galaxies and {\it Planck} CMB lensing \citep{2022JCAP...02..007W}. Recently, however, the analysis \citep{2023arXiv230405203M} using the high-precision CMB lensing measurement from the Atacama Cosmology Telescope DR6 measured $\sigma_8$ consistent with that of the {\it Planck} $\Lambda$CDM model, i.e. no $\sigma_8$ tension. 
\citet{2023arXiv230905659F} later used the tomographic cross-correlations of the ACT lensing and the unWISE galaxies, confirming no $\sigma_8$ tension, although the redshift distribution of the unWISE galaxies is not yet certain and the analysis used the different theoretical model from that of \citep{2022JCAP...02..007W}.
Thus, if these different results are of physical origin, and not due to observational systematic effects, we need to modify the $\Lambda$CDM model over different wavelength scales and different redshift windows in 
a non-trivial way. This is an interesting direction to explore. 

There are promising observational avenues for exploring baryonic feedback effects on large-scale structure. 
Cross-correlation methods of the thermal and kinetic Sunyaev-Zel'dovich (SZ) effects \citep[][]{2021PhRvD.103f3513S,2023MNRAS.525.1779P,2024arXiv240200110T} and the fast radio burst (FRB) \citep{2022ApJ...928....9L} with foreground galaxy distribution would be powerful tools to extract the baryon distribution around the host halos of the galaxies (the thermal electrons for the thermal SZ effect and the free electrons for the kinetic SZ and FRB, respectively). Improving the precision of measurements of the baryon fraction in galaxy clusters, by combining weak lensing with the thermal SZ and X-ray observations for a cluster sample, would also be useful to infer the amount of AGN feedback effects in the cluster environment, which is most relevant to the scales ($k\sim 1~h{\rm Mpc}^{-1}$) of weak lensing cosmology \citep[see Fig.~23 in Ref.][]{FLAMINGO_project}. 

The HSC~Year3 data we used in this paper is still an intermediate dataset of the Subaru HSC Survey, and 
we will have about a factor of 2.5 larger (about 1,100~sq. degrees) data for the final-year HSC data. 
We therefore expect an improvement in the weak lensing cosmology and also in the constraint on the baryonic effects in the cosmic shear signal using the method developed in this paper. In addition, the Subaru 
Prime Focus Spectrograph \citep[][]{2014PASJ...66R...1T} will soon carry out a spectroscopic survey of (emission-line) galaxies in the same large-scale structures as in the HSC galaxies,  
over the same area of the sky and at the same Subaru Telescope. Combining the imaging and spectroscopic data for the same sky, as well as other datasets such as CMB, 
will promise to improve the cosmological constraints and to constrain the baryonic effects in structure formation.

\begin{acknowledgments}
R.~T. and M.~T. thank Kaz~Akitsu, Nick~Kokron and Steven~Chen for useful discussion. 
This work was supported in part by World Premier International Research Center Initiative (WPI Initiative), MEXT, Japan, 
JSPS KAKENHI Grant Numbers JP19H00677, 
JP20H05850, JP20H05855, JP20H05861, JP21H01081, JP22H00130, JP22K03634, JP22K21349, JP23H00108, JP23H04005, JP23KJ0747, and 
by Basic Research Grant (Super AI) of Institute for AI and Beyond of the University of Tokyo. 
{
H.~M is supported by Tokai Pathways to Global Excellence (T-GEx), part of MEXT Strategic Professional Development Program for Young Researchers.}
This work was in part performed 
at the Center for Data-Driven Discovery, Kavli IPMU. We also acknowledge financial support from Japan Science and Technology Agency (JST) AIP Acceleration Research Grant Number JP20317829. Numerical simulations for the learning data set used to build our emulator were carried out on Cray XC50 at the Center for Computational Astrophysics, National Astronomical Observatory of Japan.

The Hyper Suprime-Cam (HSC) collaboration includes the astronomical communities of Japan and Taiwan, and Princeton University. The HSC instrumentation and software were developed by the National Astronomical Observatory of Japan (NAOJ), the Kavli Institute for the Physics and Mathematics of the Universe (Kavli IPMU), the University of Tokyo, the High Energy Accelerator Research Organization (KEK), the Academia Sinica Institute for Astronomy and Astrophysics in Taiwan (ASIAA), and Princeton University. Funding was contributed by the FIRST program from Japanese Cabinet Office, the Ministry of Education, Culture, Sports, Science and Technology (MEXT), the Japan Society for the Promotion of Science (JSPS), Japan Science and Technology Agency (JST), the Toray Science Foundation, NAOJ, Kavli IPMU, KEK, ASIAA, and Princeton University. This paper makes use of software developed for the Large Synoptic Survey Telescope. We thank the LSST Project for making their code available as free software at \url{http://dm.lsst.org}

The Pan-STARRS1 Surveys (PS1) have been made possible through contributions of the Institute for Astronomy, the University of Hawaii, the Pan-STARRS Project Office, the Max-Planck Society and its participating institutes, the Max Planck Institute for Astronomy, Heidelberg and the Max Planck Institute for Extraterrestrial Physics, Garching, The Johns Hopkins University, Durham University, the University of Edinburgh, Queen's University Belfast, the Harvard-Smithsonian Center for Astrophysics, the Las Cumbres Observatory Global Telescope Network Incorporated, the National Central University of Taiwan, the Space Telescope Science Institute, the National Aeronautics and Space Administration under Grant No. NNX08AR22G issued through the Planetary Science Division of the NASA Science Mission Directorate, the National Science Foundation under Grant No. AST-1238877, the University of Maryland, and Eotvos Lorand University (ELTE) and the Los Alamos National Laboratory.

Based in part on data collected at the Subaru Telescope and retrieved from the HSC data archive system, which is operated by Subaru Telescope and Astronomy Data Center at National Astronomical Observatory of Japan.
\end{acknowledgments}

\twocolumngrid

\bibliography{refs}

\end{document}